\documentclass{aa}  

\usepackage{graphicx}
\usepackage{txfonts}
\PassOptionsToPackage{hyphens}{url}\usepackage{hyperref}
    \makeatletter
    \renewcommand*\aa@pageof{, page \thepage{} 
        of \pageref*{LastPage}}
    \makeatother
\usepackage{paralist} 
\usepackage{xspace}
\usepackage{url}
\usepackage{textcomp}
\usepackage{numprint}
\usepackage{flushend} 

\def \acc {0.984}
\def \tltr {0.991}
\def \tetr {0.90}
\def \eg {\emph{e.g.,\xspace}}

\bibpunct{(}{)}{;}{a}{}{,} 

\def \so {\textsubscript{OTELO}}
\def \sdnn {\textsubscript{DNN}}
\def \ngal {370\xspace}
\def \ntrain {192\xspace}   
    \def \npure {128}       
    \def \nval {64\xspace}  
\def \ntest {178\xspace}    
\def \bl {0.92\xspace}

\begin{document}

   
  \title{Non-Sequential Neural Network for Simultaneous, Consistent Classification and Photometric Redshifts of OTELO Galaxies}


\author{
   		  \mbox{José A. de Diego 
   		  \inst{\ref{inst:unam}}}
   		\and
          \mbox{Jakub Nadolny 
          \inst{\ref{inst:iac}, \ref{inst:ull}}}
        \and
          \mbox{Ángel Bongiovanni 
          \inst{\ref{inst:iram}, \ref{inst:aspid}}}
        \and
          \mbox{Jordi Cepa 
          \inst{\ref{inst:iac}, \ref{inst:ull}, \ref{inst:aspid}}}
        \and
          \mbox{Maritza A. Lara-López 
          \inst{\ref{inst:armagh}}}
        \and
          \mbox{Jesús Gallego 
          \inst{\ref{inst:ucm}}}
        \and
          \mbox{Miguel Cerviño 
          \inst{\ref{inst:inta}}}
        \and
          \mbox{Miguel Sánchez-Portal 
          \inst{\ref{inst:iram}, \ref{inst:aspid}}}
        \and
          \mbox{J. Ignacio González-Serrano 
          \inst{\ref{inst:aspid}, \ref{inst:ifca}}}
        \and
          \mbox{Emilio J. Alfaro 
          \inst{\ref{inst:iaa}}}
        \and
          \mbox{Mirjana Povi\'{c} 
          \inst{\, \ref{inst:essti}, \ref{inst:iaa}}}
        \and
          \mbox{Ana María Pérez García 
          \inst{\ref{inst:aspid}, \ref{inst:inta}}}
        \and
          \mbox{Ricardo Pérez Martínez 
          \inst{\ref{inst:aspid}, \ref{inst:esa}}}
        \and
          \mbox{Carmen P. Padilla Torres 
          \inst{\ref{inst:iac}, \ref{inst:ull}, \ref{inst:inaf}, \ref{inst:cefca}}}
        \and
          \mbox{Bernabé Cedrés
          \inst{\ref{inst:iac}, \ref{inst:ull}}}
        \and
          \mbox{Diego García-Aguilar
          \inst{\ref{inst:unam}}}
        \and
          \mbox{J. Jesús González \inst{\ref{inst:unam}}}
        \and
          \mbox{Mauro González-Otero
          \inst{\ref{inst:iac}, \ref{inst:ull}}}
        \and
          \mbox{Rocío Navarro-Martínez
          \inst{\ref{inst:aspid}}}
        \and
          \mbox{Irene Pintos-Castro 
          \inst{\ref{inst:cefca}, \ref{inst:toronto}}}
          }

   \institute{
   Instituto de Astronomía,
   	Universidad Nacional Autónoma de México,
    Apdo. Postal 70-264, 04510 Ciudad de México, Mexico\\
    \email{jdo@astro.unam.mx} \label{inst:unam}
    		  \and
   Instituto de Astrofísica de Canarias (IAC),
    E-38200 La Laguna, Tenerife, Spain \label{inst:iac}
              \and
   Departamento de Astrofísica,
    Universidad de La Laguna (ULL), 
    E-38205 La Laguna, Tenerife, Spain \label{inst:ull}
              \and
   Institut de Radioastronomie Millimétrique (IRAM),
    Av. Divina Pastora 7, Local 20, 
    18012 Granada, España \label{inst:iram}
              \and
   Asociación Astrofísica para la Promoción de la Investigación,
    Instrumentación y su Desarrollo, ASPID, 
    E-38205 La Laguna, Tenerife, Spain \label{inst:aspid}
              \and
   Armagh Observatory and Planetarium, 
    College Hill, 
    Armagh, BT61 DG, UK \label{inst:armagh}
              \and
   Departamento de Física de la Tierra y Astrofísica. 
    Instituto de Física de Partículas y del Cosmos (IPARCOS).
    Universidad Complutense de Madrid, 
    E-28040 Madrid, Spain \label{inst:ucm}
              \and
   Depto. Astrofísica, 
   	Centro de Astrobiología (INTA-CSIC),
   	ESAC Campus, Camino Bajo del Castillo s/n, 
   	28692, Villanueva de la Cañada, Spain \label{inst:inta}
              \and
   Instituto de Física de Cantabria
    (CSIC-Universidad de Cantabria), 
    E-39005 Santander, Spain \label{inst:ifca}
              \and
   Instituto de Astrofísica de Andalucía, CSIC,
    Glorieta de la Astronomía, s/n, 
    18008 Granada, Spain\label{inst:iaa}
              \and
   Ethiopian Space Science and Technology Institute  
    (ESSTI),
    Entoto Observatory and Research Center (EORC), 
    Astronomy and Astrophysics Research and Development Department, 
    PO Box 33679, Addis Ababa, Ethiopia \label{inst:essti}
              \and
   ISDEFE for European Space Astronomy Centre (ESAC)/ESA,
    P.O. Box 78, 
    E-28690, Villanueva de la Cañada, 
    Madrid, Spain \label{inst:esa}
             \and
   Fundación Galileo Galilei, 
   	Telescopio Nazionale Galileo, 
   	Rambla José Ana Fernández Pérez, 7, 
   	38712 Breña Baja, Santa Cruz de la Palma, Spain \label{inst:inaf}
              \and
   Centro de Estudios de Física del Cosmos de Aragón (CEFCA), 
    Plaza San Juan, 1, E-44001, Teruel, Spain \label{inst:cefca}
              \and
   Department of Astronomy \& Astrophysics,
    University of Toronto, Canada \label{inst:toronto}
             }

	\date{Version: \today}

 
  \abstract
   { 
    Computational techniques are essential for mining large databases produced in modern surveys with value-added products.}
   { 
    This paper presents a machine learning procedure to carry out simultaneously galaxy morphological classification and photometric redshift estimates.   Currently, only spectral energy distribution (SED) fitting has been used to obtain these results all at once.}
   { 
    We used the ancillary data gathered in the OTELO catalog and designed a non-sequential neural network that accepts optical and near-infrared photometry as input.  The network transfers the results of the morphological classification task to the redshift fitting process to ensure consistency between both procedures.}
   { 
    The results successfully recover the morphological classification and the redshifts of the test sample, reducing catastrophic redshift outliers produced by SED fitting and avoiding possible discrepancies between independent classification and redshift estimates.  Our technique may be adapted to include galaxy images to improve the classification.}
   {}


\titlerunning{ Non-sequential Neural Network for OTELO Galaxies}

   \maketitle
%

\section{Introduction}


Massive morphological classification and photometric redshift estimates of galaxies are principal drivers of the ongoing and future imaging sky surveys because they are necessary to understand galaxy formation, evolution, physical and environmental properties, and constraining cosmological models.
%
%
Classification of galaxies often relies on their shape determined by visual inspection, providing (usually) the following classes: spiral, elliptical, lenticular, and irregular.  However, visual classification is a task that requires an enormous amount of telescope and human time, and the results depend on the experience of the classifier.  Thus, this technique is unfeasible for a large number of small images of dim galaxies, as expected in current and future extensive surveys.  Fortunately, machine learning techniques are handy for this task, given that we can gather enough training data of classified galaxies.  This problem inspired the Galaxy Zoo survey, derived from a vast citizen science initiative to assist galaxies' morphological classification.  The first version comprised over \numprint{900000} SDSS galaxies brighter than $r = 17.7$ \citep{lintott11}; successive additions to the catalog include additional SDSS and Hubble data \citep{willett13, willett17}, and the CANDELS survey \citep{simmons17}.

Spectroscopic observations provide an accurate answer to classification and redshift problems, but it is also a time-demanding solution, limited by the spectral range of multi-object spectrographs and impractical for the dimmest galaxies.  
Moreover, spectroscopic surveys may be susceptible to several selection biases, some of which can potentially affect the representation of different galaxy morphological types \citep{cunha14}.  Therefore,  studies routinely rely on photometric broadband data to investigate large-scale structures and galaxy evolution.  By low-resolution sampling of the spectral energy distribution (SED), it is possible to discriminate the nature of a source (galaxy or star), render a morphological SED-based classification, and provide a redshift estimate.

A popular technique to achieve both morphological classification and photometric redshift estimate consists of fitting empirical or theoretical SED templates
\citep[\eg][]{kinney96, bruzual03, ilbert09} to the broadband data.  However, this technique is complex because of the observed SED's dependence on the contribution of intense emission features, absorption by the galaxy's gas and the Milky Way, dust attenuation, and optical system and Earth's atmosphere transmission.  These effects, along with some large errors, may contribute to producing a significant number of outliers compared with spectroscopic techniques.

The SDSS has offered the opportunity to introduce machine learning algorithms to analyze photometric data and broadband images of large samples of extragalactic objects \citep[\eg][]{huertas:2008, banerji10, carrasco13, dieleman:2015, tuccillo15, sanchez2018}.  Additionally, the high-resolution images achieved by the Hubble Space Telescope (HST) have encouraged the study of the automatized determination of morphological parameters and classification \citep[\eg][]{odewahn96, bom17, pourrahmani18, tuccillo18}.  These algorithms will be essential for dealing with the bulk of data that next-generation surveys, such as LSST and Euclid, will gather during the oncoming years.

Machine learning aims to make accurate predictions after learning how to map input data to outputs.  The most popular machine learning techniques employed in Astronomy either for morphological classification ---we will refer to as just classification--- or redshift estimates are Support Vector Machines \citep[SVM; \eg][]{wadadekar2005, jones17, khramtsov2020} and particularly the GalSVM code \citep{huertas:2008, povic12, povic13, povic15, pintos16, amado19}, Random Forests \citep[\eg][]{carrasco13, mucesh21}, and neural networks \citep[\eg][]{serra93, serra96, firth03, sanchez2018, diego20}.  All these are supervised techniques; they need a labeled dataset for training.  SVMs map the data space into an extra dimension that maximizes the separation between categories or best correlates with the regression target variable.  Random Forests are sets of decision trees, each performing a separate classification or regression task; the output of a Random Forest is the mode of the classes or the average of the decision trees.  A neural network is a set of interconnected nodes or neurons that loosely reproduces brain performance.  Each machine learning method has its strengths and weaknesses related to its interpretability (associated with the number of parameters and nonlinear approximations), the amount of training data required, and the optimization algorithm dependence.

Neural networks have several characteristics that probably convert them into the handiest machine learning technique to address future astronomical survey analysis.  In contrast to other methods, neural networks need little, if any, feature engineering; after coding non-numerical records, and normalization or standardization, the network can analyze the data without further manipulation.  It is even possible to combine data in different formats, such as tabulated records, images, and spectra.  Besides, neural networks can update learning by further training on additional datasets.  Another quirk that makes them very versatile is transfer-learning, which consists of applying pre-trained neural networks for a particular task to different assignments.  Thus, neural networks trained using datasets containing millions of heterogeneous images (\eg\ animals, vehicles, plants, tools) such as ImageNet \citep{deng2009} may transfer their capacity to recognize image profiles to databases from domains such as medical imaging or geology \citep[\eg][]{marmanis2016, raghu2019}.  Recently, \citet{eriksen20} applied transfer-learning from a network trained with photometric data simulations to obtain photometric redshifts for galaxies from the Physics of the Accelerating Universe Survey \citep{padilla19}.

Previous neural networks used in morphological classification and redshift estimates of galaxies are of the sequential type.  Such neural networks consist of a neat stack of layers with exactly one input and one output tensor per layer.  
%
Sequential neural networks are the most common kind of architecture, but they cannot deal with complicated graph topologies such as multi-branch models, layer sharing, or multiple inputs and outputs.  Consequently, machine learning solutions handle morphological classification and redshift estimates employing different models, unlike SED fitting codes.  For example, \citet{firth03} use different neural networks for these two tasks, but they do not compare whether the classifications and the redshifts are congruent or not.  In contrast, several authors report degenerate solutions when using templates to fit the SED.  For instance, \citet{raihan20} report catastrophic redshift outliers for galaxies with spectral redshifts at \mbox{$2 < z_{spec} <3$} wrongly assigned at redshifts below 0.3, which they attribute to incorrect template matching and degeneracies of the color-redshift dependency. Besides, \citet{zhang09} find that low redshift dusty edge-on spirals may be misclassified as early-type galaxies because of their similar $u-r$ color.  Also, \citet{salvato19} report that the $i-z$ color could correspond to multiple redshifts.  Though the use of multiple photometric bands comprising a broad range of wavelengths can alleviate the color degeneracy \citep{benitez09}, it is still possible to obtain inconsistent results when using different classification and redshift procedures.  

This paper presents the use of a non-sequential, multi-output Deep Neural Network (DNN) to yield both galaxy morphological SED-based classification and redshift estimates, with minimal data handling of optical and infrared photometry.  Furthermore, the network model allows for feeding back the galaxy classification result to ensure internal consistency with the photometric redshift estimate.

\begin{figure}[t]
    \centering
    \includegraphics[width=\columnwidth]{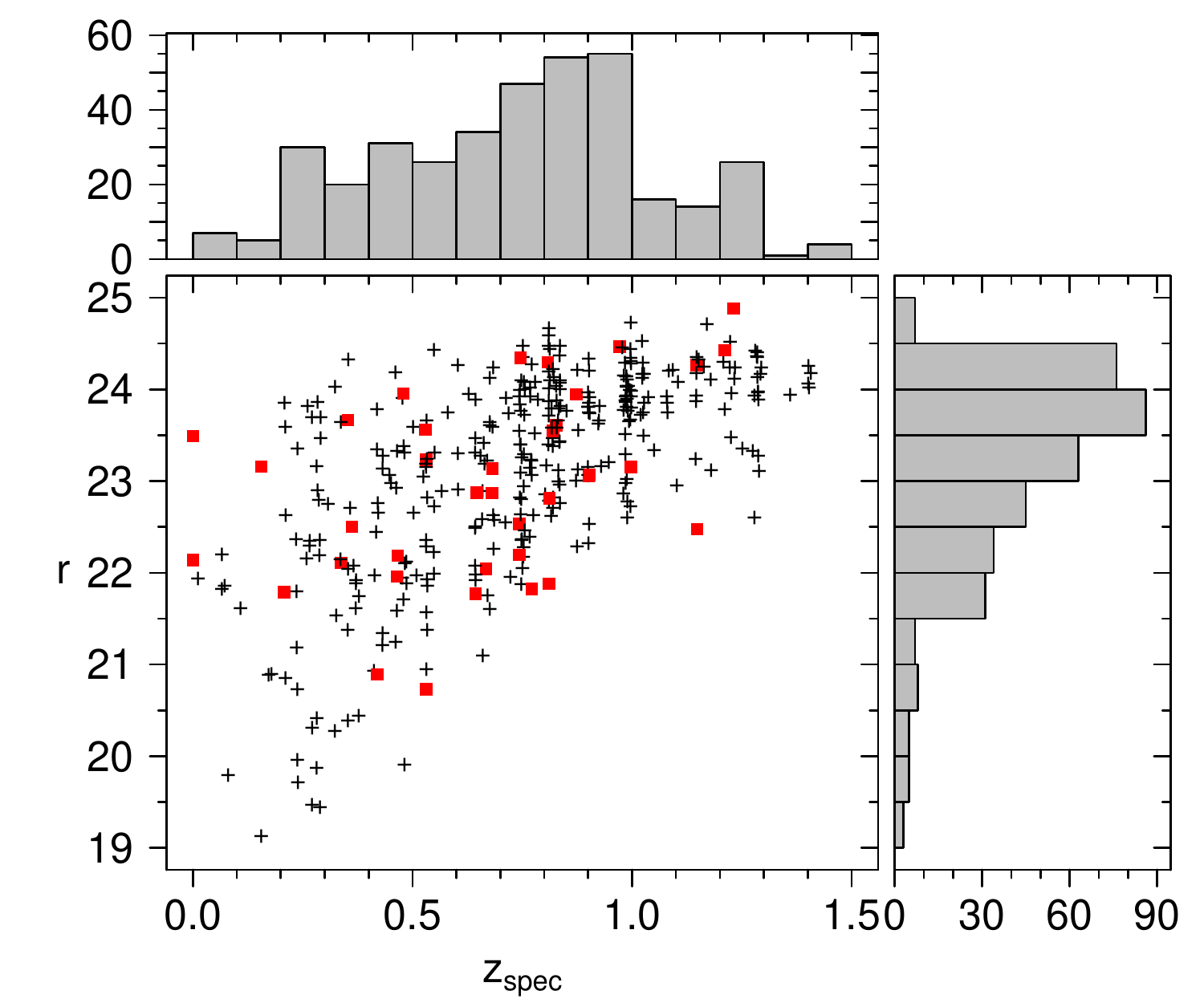}
        \caption{Data distribution of the sample used in this work.  Bottom-left panel scatter plot shows the $r$-magnitude vs. spectral redshift distribution, where black crosses represent LT galaxies and red boxes ET. Bottom-right panel histogram indicates the $r$-magnitude distribution. Upper panel histogram displays the spectral redshift distribution of the sample.}
    \label{fig:distribution}
\end{figure}

\section{Methods}



\subsection{Photometric data}

This paper presents a multi-output DNN to classify galaxies as early or late types and estimate their redshifts, using AB-system photometric data of a sample extracted from the \href{http://research.iac.es/proyecto/otelo}{OTELO catalog}\footnote{\url{http://research.iac.es/proyecto/otelo}} \citep{bongiovanni19}.  OTELO is a deep survey carried out with the OSIRIS Tunable Filters instrument attached at the 10.4~m Gran Telescopio Canarias (GTC). The OTELO catalog includes OTELO detections and reprocessed ancillary data from Chandra, GALEX, CFHT, WIRDS, and IRAC covering from ultraviolet to the mid-infrared.  Our sample consists of \ngal OTELO galaxies included in the \citet{nadolny21} morphological analysis, of which 31 are early-type and 339 late-type.  All these galaxies have spectral redshift $z < 1.41$ obtained from the DEEP2 Galaxy Redshift Survey \citep{newman13, bongiovanni19}.  These redshifts have high-quality parameters $Q \geq 3$ that indicate that they are very reliable \citep{newman13}.  The galaxies also have $ugriz$ optical photometry from the \href{http://www.cfht.hawaii.edu/Science/CFHTLS/}{Canada-France-Hawaii Telescope Legacy Survey}\footnote{\url{http://www.cfht.hawaii.edu/Science/CFHTLS/}} (CFHTLS), and $JHKs$ near-infrared photometry obtained from the \href{https://www.cadc-ccda.hia-iha.nrc-cnrc.gc.ca/en/ cfht/wirds.html}{WIRcam Deep Survey}\footnote{\url{https://www.cadc-ccda.hia-iha.nrc-cnrc.gc.ca/en/ cfht/wirds.html}} (WIRDS); the dataset does not contain missing values.

Figure~\ref{fig:distribution} shows the $r$-band magnitudes vs. the redshift, distinguishing between early-type (ET) and late-type (LT) galaxies.  The $r$-magnitude distribution expands from $r=19$ to $r=25$. Table~\ref{tab:stat} presents the medians and interquartile ranges of the $r$ magnitudes and the spectroscopic redshifts, discriminating between ET and LT galaxies and the total sample.  Neither the Wilcoxon test for differences between medians nor the Kolmogorov-Smirnov probe for differences between distributions yielded statistically significant dissimilarities in the $r$ magnitudes or the spectral redshifts between the two types of galaxies.  We note that this is a bright sample for OTELO's sensitivity, as the OTELO survey attains completeness of 50\% at $r=26.38$ \citep{bongiovanni19}.  The $r$-magnitude distribution is left-skewed ($\text{skewness} = -1.1$), with a sharp decay indicating a strong deficiency of dim sources.  Therefore, this is not a complete sample and should not be used for cosmological inferences.  For the redshifts, they expand from $z=0$ up to $z \simeq 1.4$.  The sample distribution shows a small number of galaxies at $z < 0.2$ and a sharp decay for redshifts $z > 1$ that is inherited from the DEEP2 Galaxy Redshift Survey.

We transformed the OTELO data to be used by the neural network.  Thus, we standardized the photometric data using the Z-Score with mean zero and standard deviation one.  The record field \texttt{MOD\_BEST\_deepN} in the OTELO catalog lists the LePhare templates \citep{arnouts99, ilbert06} that best fit the SED to ancillary data, coded from 1 to 10  \citep{bongiovanni19}.  We use this information to build a morphological SED-based classification (or just classification from now on).  The best fit for galaxies coded as "1" is the E/S0 template from \citet{coleman80}, and we coded them as ET in our sample.  The best fit for galaxies coded from 2 to 10 corresponds to late-type galaxies templates from \citeauthor{coleman80} and starbursts from \citet{kinney96}, and we assigned the LT code to all of them.

\begin{table}[t]
\caption{Data statistics}\label{tab:stat} 
\centering
\begin{tabular}{lcccc}
  \hline\hline
 & $\tilde{r}$ & IQR$_r$ & $\tilde{z}_{spec}$ & IQR$_Z$ \\ 
  \hline
ET & 23.13 & 1.46 & 0.772 & 0.256 \\ 
  LT & 23.39 & 1.42 & 0.755 & 0.503 \\ 
  Total & 23.35 & 1.45 & 0.755 & 0.494 \\ 
   \hline
\end{tabular}
\end{table}

\subsection{Neural network}

Neural networks are function approximation algorithms.  The main assumption is that there is an unknown function $f$ that maps input data to output data.  The Universal Approximation Theorem \citep{cybenko89, hornik91, leshno93} implies that ---no matter how complex the function $f$ may be--- there is a neural network that can approximate the function with arbitrary, finite accuracy.  The researcher adjusts the network architecture and hyperparameters to obtain accurate predictions.  To accomplish this task, we used two datasets: the training set for fitting the network and the testing set to provide an unbiased evaluation of the model.  Both training and testing datasets consist of available observations and labels.  Inferring the latter becomes the goal of the neural networks.  Labels may correspond to a classification scheme or a continuous unobserved variable that we want to predict. The network presented here will provide a binary classification of galaxies separated into early and late types and estimate the galaxy redshifts.



We built the neural network architecture using the functional Application Programming Interface (API) included in the Keras library for deep learning.  Keras is a high-level neural network application programming interface (API) under GitHub license, which may run under Python and R code \citep{chollet17a,chollet17b}.  With the Keras functional API, it is possible to build complex network architectures, such as multi-input and multi-output models or other non-sequentially connected layers.

In the rest of this section, we first present the analysis to build a suitable network architecture.  This analysis includes the arrangement of the training and test sets, the model description, and the comparison with some alternative architectures.  Finally, we present the model validation procedure that we will use later to present the results.

\begin{figure}[t]
    \centering
    \includegraphics[height=0.5\paperheight]{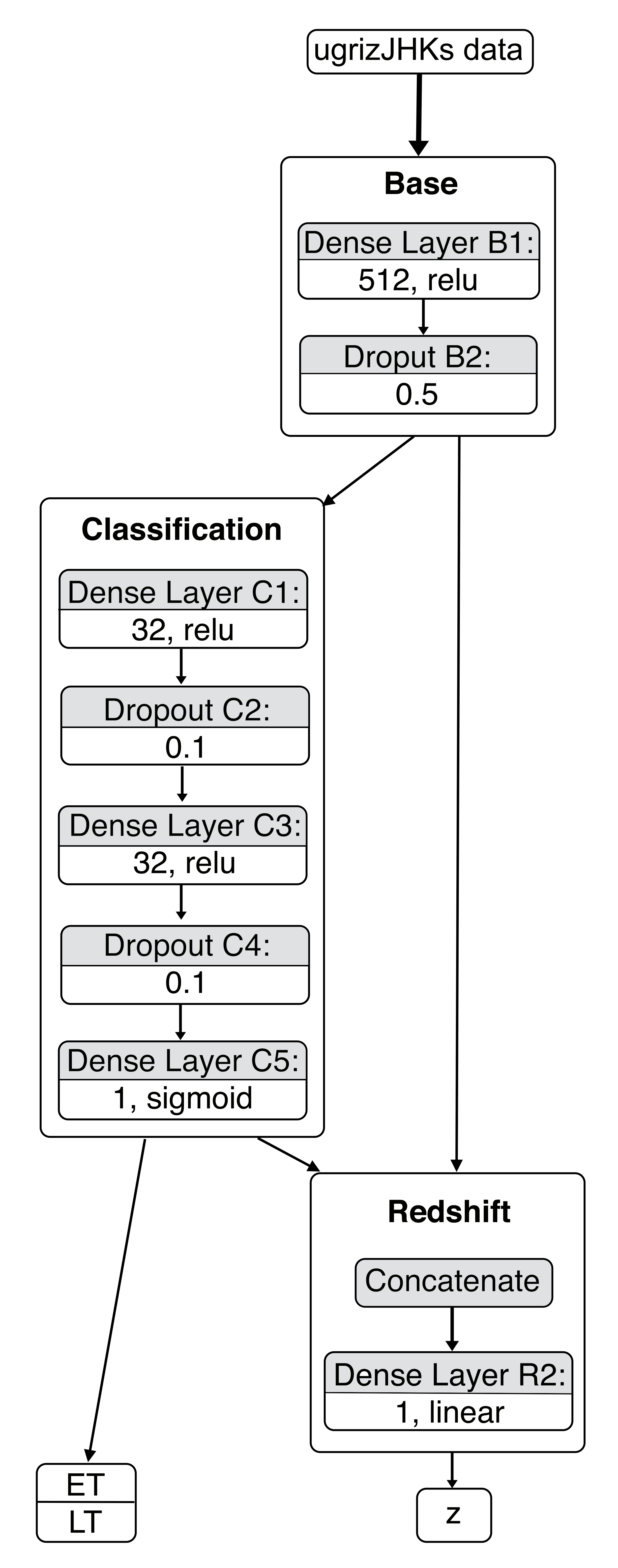}
    \caption[]{Multi-output DNN graph.  The neural architecture consists of three modules, indicated by boxes,  each comprising some layers.  The photometric input data (above) feeds the Base module, which performs common operations.  The Classification module receives its input from the Base module and outputs a binary classification probability $p$ for ET ($p_{ET} \sim 0$) and LT ($p_{LT} \sim 1$) galaxies.  The Redshift module receives inputs from the Base and the Classification modules, and outputs an estimate of the redshift. Inside each module, the individual layers are identified along with the number of neurons and the activation function (dense layers) or the dropout fraction (dropout layers).}\label{fig:dnn}
\end{figure}

\subsubsection{Determining the neural network architecture}

We pre-trained several network architectures using the pure training set while probing its performance using the validation set.  We manually added layers and neurons for this task and played with dropouts until we found no classification or redshift estimation improvement, in a similar way to previous studies \citep[\eg][]{firth03, vanzella04}. Then, we chose a set of network models that best fitted our data.  For each of these models, we performed 100 bootstrap trials, redefining the training and test sets for each trial.  These bootstrap trials provided several metric distributions that we compared to choose the final model. We will show these results later. Meanwhile, we focus on the architecture that showed the best performance.

Each training set comprises  \ntrain records, and each test dataset includes our sample's remaining \ntest galaxies.  In turn, we separated the \ntrain galaxies of the training set into two groups, the pure-training set (\npure\ galaxies) used to fit network trials and the validation set (\nval galaxies) to provide an unbiased evaluation of the fit while tuning the network hyperparameters.  The number of elements assigned to each set is a compromise: a sizeable training set improves the network fitting, while a large test set provides better statistics of its performance.  
Besides, to achieve percentual statistic precisions better than 1\%, we need a test sample larger than 100 cases.  Finally, the numbers of galaxies in the pure training and validation sets enable the processing of complete batches of 64 galaxies, optimizing the training computation.

\begin{table*}[t]
    \centering
    \caption{Bootstrap validation metric medians for some tested architectures.}\label{tab:performance}
    \begin{tabular}{l r@{$\,\pm\,$}l r@{$\,\pm\,$}l r@{$\,\pm\,$}l r@{$\,\pm\,$}l}
        \hline\hline
        Metric & \multicolumn{2}{c}{Arch. 1} & \multicolumn{2}{c}{Arch. 2} & \multicolumn{2}{c}{Arch. 3} & \multicolumn{2}{c}{Arch. 4} \\
        \hline
        \rule{0pt}{1.2\normalbaselineskip}
        REDSHIFT \\
        \hspace{1em} Mean square error                          & 0.10 & 0.03 & 0.10 & 0.03 & 0.10 & 0.03 & 0.11 & 0.02  \\ 
        \hspace{1em} Mean absolute error                        & 0.35 & 0.03 & 0.36 & 0.04 & 0.34 & 0.03 & 0.35 & 0.03 \\[5pt] 
        CLASSIFICATION \\
        \hspace{1em} Binary cross-entropy.                      & 0.04 & 0.03 & 0.04 & 0.03 & 0.04 & 0.03 & 0.04 & 0.04 \\ 
        \hspace{1em} Area under the curve                       & 0.99 & 0.02 & 0.99 & 0.01 & 0.99 & 0.01 & 0.99 & 0.01 \\[5pt] 
        TOTAL LOSS                                              & 0.13 & 0.02 & 0.15 & 0.04 & 0.14 & 0.03 & 0.14 & 0.04 \\ 
        \hline
    \end{tabular}
\end{table*}

Figure~\ref{fig:dnn} shows the architecture of our final neural network.  The input is the photometric data, which feeds the Base module of layers.  The first element in this module consists of a dense layer formed by 512 nodes with a Rectified Linear Unit (ReLU) activation function, defined as the positive part of its argument: $f_{ReLU}(x) = \max (0,x)$.  ReLU's main advantage over other activation functions is that it alleviates the vanishing gradient problem compared to other activations ---especially those with a limited output range--- such as the sigmoid and the hyperbolic tangent. Besides, ReLU computation is fast and produces sparse activation ---only a fraction of the hidden neurons will activate.  The last piece in the first module is a dropout layer that randomly resets half of the previous layer weights to avoid overfitting.

The output of the Base module feeds both the Classification and the Redshift branches.  The Classification module comprises five layers, the first four alternating a dense layer with 32 nodes and a ReLU activation function, and a dropout which resets 10\% of the previous weights.  The last layer in this module consists of a single node with a sigmoid activation defined as $f_{\mathrm{sigmoid}}(x) = \frac{1}{1+e^{-x}}$ that yields a real number between 0 and 1. If employed at an output layer, the sigmoid directly maps into a binary classification probability such as early or late-type galaxies, in our case.  This result constitutes one of the network outputs and also feeds the Redshift module.

The ET and LT galaxy classification proved to be very reliable for z < 2 \citep{diego20} and provides a continuous probability rather than a simple binary output.  Therefore, we decided to transfer the classification probability to the Redshift module to ensure consistency between both procedures.  This module then receives two inputs delivered by the Base and the Classification elements.  The Redshift module's first piece is a concatenate layer that combines the two inputs and passes the information to a single node layer with linear activation ($f_{linear}(x) = a x$) that provides the estimated redshift output.  The linear activation has a constant derivative ($f^\prime_{linear} = a$) that prevents gradient optimization.  Hence, linear activation is only useful for simple tasks aiming at easy interpretability, and in practice, they are used only at output layers to provide a regression value.

We use the following notation to describe our model.  We code a dense layer as $D^n$, where $n$ is the number of neurons. If the next element is a dropout layer, we denote the dense and dropout set as $D^n_f$, where $f$ is the dropout fraction.  Similarly, we code concatenation layers as $C^y_x$, where the letter $y$ states for the Base ($b$) or the Classification ($c$) modules, and the letter $x$ for the Classification module, or the previous layer ($+$, which we will use when discussing alternative architectures).  With this notation, the modules of our network become:

\begin{itemize}
    \item Base: $D^{512}_{0.5}$
    \item Classification: $D^{32}_{0.1}:D^{32}_{0.1}:D^1$
    \item Redshift: $C^b_c:D^1$
\end{itemize}


Apart from the network architecture, the model needs a set of control hyperparameters to improve performance, which may concern the interested reader.  
%
Thus, we used two loss functions included in Keras.  For classification involving two groups (early and late-type galaxies), the Binary Cross-Entropy  (or log loss) is the fiducial loss function.  We chose the Mean Square Error for redshift estimates, a standard loss function for regression problems.  The network uses a Root Mean Square Propagation (RMSprop) optimizer for quickly minimizing the loss functions.  RMSprop is a popular adaptive learning rate optimizer for neural networks that allows splitting a training sample in mini-batches ---typically of sizes 32, 64, or 128 elements each. Splitting the training sample in mini-batches improves the computer memory usage and speeds the calculus.

Finally, while defining the network architecture, we added two Keras callback functions that perform actions to control the training.  The `EarlyStopping' callback stops training when the validation loss metric for galaxy classification has not improved after a specific number of training epochs, in our case 10.  The `ReduceLROnPlateau' callback reduces the optimizer's learning rate by a given factor when the validation loss metric for redshift estimate has stopped improving after a certain number of training epochs; we reduced the learning rate by a factor of 0.9, and we set the number of epochs in 5.  We kept the resulting number of training epochs and fitted a sigmoid function to the evolution of the learning rate as a function of the training epochs.  We used these parameters to obtain the final results ---when no validation set was available--- fixing the number of training epochs and using the Keras `LearningRateScheduler` callback to control the training.

Table~\ref{tab:performance} shows the metric medians for 100 bootstrap trials using some of the network architectures that we have tested.  We show the median statistic instead of the mean because some of the metrics present skewed distributions.  Architecture~1 corresponds to the final model that we have discussed so far.  Architectures~2, 3, and 4 are different versions obtained by modifying the modules Base, Classification, and Redshift, respectively, and maintaining the other modules unchanged, as indicated below:

\begin{itemize}
    \item Arch. 2 – Base: $D^{64}_{0.5}$
    \item Arch. 3 – Classification: $D^{32}_{0.1}:D^1$
    \item Arch. 4 - Redshift: $D^{16}:C^c_+:D^1$
\end{itemize}

Table~\ref{tab:performance} provides an overview of the architecture performance dependency for moderate changes in each module.  Statistically, all the architectures show similar performances, indicating that the results are remarkably stable for mild architecture variations.  Nevertheless, the Total loss for Architecture~1 achieves the best overall performance.  The Total loss is the sum of the loss functions used for optimization (the Mean square error and the Binary cross-entropy).

\subsubsection{Leave-One-Out Cross-Validation}

Once we had obtained a network model that achieved a good performance, we ran a Leave-One-Out Cross-Validation \citep[LOOCV;][]{gareth13}. LOOCV is a specific implementation of the K-fold Cross-Validation, which consists of training using the whole labeled dataset but leaving one element out for testing.  Then we repeat this procedure for each component of the dataset.   In our case, the LOOCV procedure involves \ngal different training runs, one for each galaxy in our sample using the other 369 as the training set.   The advantage of LOOCV is that all the records contribute to both the training and test sets, which are the largest we can achieve, given the available data.  Thus LOOCV is especially helpful for evaluating the performance of small datasets like ours.  We present the results obtained using the LOOCV procedure in the next section.




\section{Results}

\subsection{Classification}

Previous studies have shown the power of colors and morphological parameters to implement automatic galaxy classification at redshifts $z < 0.4$.  Thus, \citet{strateva01} using the $u-r$ discriminant color achieve an accuracy ---the ratio between the number of correctly classified cases and the total number of cases--- of 0.72, and \citet{deng13} using concentration indexes reaches an accuracy of 0.96.  \citet{vika15} employing the $u-r$ color discriminant and the Sérsic index get an accuracy of 0.89.  Recently, \citet{diego20} used optical and near-infrared photometry along with either the Sérsic index or the C$_{80/20}$ concentration index to train sequential neural networks achieving accuracies greater than 0.98.

As stated above, the classification module of our neural network ends with a sigmoid activation function layer.  This function yields a probability between zero and one for each galaxy.  Near-zero probabilities correspond to ET classified galaxies, while near-one probabilities imply LT classification.  We fix the threshold probability at $p_{th} = 0.5$, which is the usual choice for classification problems.  Thus, the neural network classifies galaxies with probabilities $p < p_{th}$ as ET, and LT otherwise.

\begin{table}[t]
    \centering
    \caption{Contingency table with default 0.5 threshold.}\label{tab:contingency}
    \begin{tabular}{lrrr}
        \hline\hline
                 & ET\so & LT\so & Sum \\ 
        \hline
        ET\sdnn  & 28    &   3   &  31 \\ 
        LT\sdnn  &  3    & 336   & 339 \\ 
        Sum      & 31    & 339   & 370 \\ 
        \hline
    \end{tabular}
\end{table}
%

\begin{table*}[t]
\centering
\caption{Misclassified galaxies.} 
\label{tab:misclassified}
\begin{tabular}{rcccrrr}
  \hline\hline
  ID & Visual & OTELO & DNN & z\_.spec. & z\_.OTELO. & z\_.DNN. \\ 
  \hline
  26   &    & ET & LT & 0.75 & 0.62 & 0.79 \\ 
  1482 & LT & LT & ET & 0.53 & 0.68 & 0.56 \\ 
  1483 & LT & ET & LT & 1.02 & 0.83 & 0.99 \\ 
  2170 & LT & LT & ET & 0.68 & 0.82 & 0.71 \\ 
  5229 &    & LT & ET & 1.15 & 0.85 & 0.98 \\ 
  6176 & LT & ET & LT & 0.32 & 0.25 & 0.42 \\ 
   \hline
\end{tabular}
\end{table*}

For classification purposes of our galaxy sample, it is convenient to define a  baseline for evaluating our method's accuracy improvement.  In this case, where the sample LT proportion ($\hat{p}_{LT} = 0.92$) dominates over the ET galaxies (($\hat{p}_{ET} = 0.08$), assigning all the galaxies to the LT class yields a baseline accuracy of 0.92
%
Table~\ref{tab:contingency} shows the contingency table for the classification results.  From the 31 ET and 339 LT OTELO galaxies, the DNN recovers 28 and 336, respectively. Three ET and three LT galaxies are mismatched.
The accuracy attained by the neural network can be inferred from this table, resulting in $\acc \pm 0.007$.  This accuracy represents a considerable improvement compared with the classification baseline of \bl.  Similarly, the True LT Rate and the True ET Rate can be gathered from this table, resulting in $\tltr \pm 0.006$ and $\tetr \pm 0.06$, respectively.  
%
The accuracy obtained with our non-sequential model using just photometric data matches our previous result reported in \citet[$0.985 \pm 0.007$]{diego20}, where we also included the Sérsic index.

\begin{center}
\begin{figure*}[!h]
    \centering
    \begin{center}
    \begin{tabular}{ c c }
        \Large ID 00026 & \Large ID 01482 \\
        \LARGE \shortstack{HST image \\ not available} & \raisebox{-.5\height}{\includegraphics[width=0.40 \textwidth, trim = 25 0 550 25, clip]{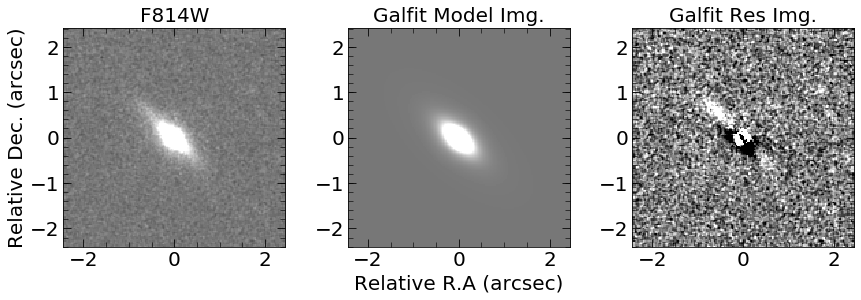}} \\

        
        \Large ID 01483 & \Large ID 02170 \\
        \raisebox{-.5\height}{\includegraphics[width=0.40 \textwidth, trim = 25 0 550 40, clip]{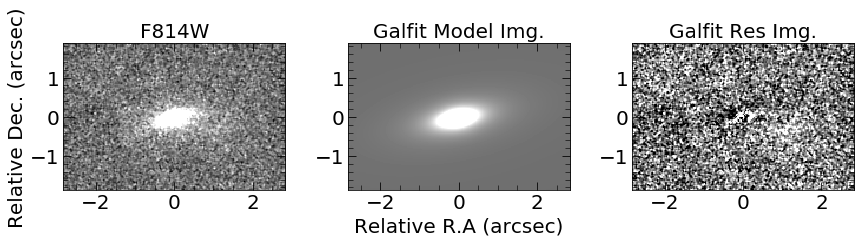}} & \raisebox{-.5\height}{\includegraphics[width=0.40 \textwidth, trim = 25 0 550 25, clip]{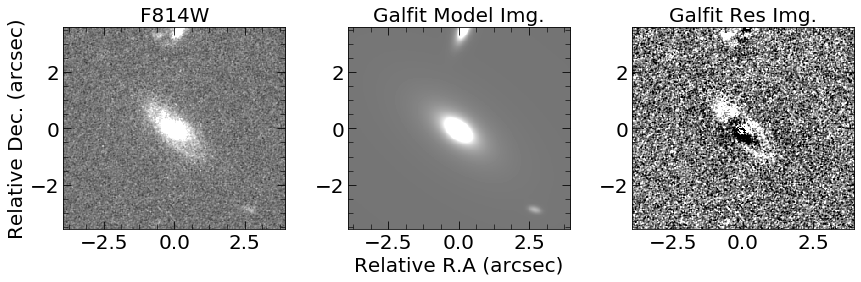}} \\ 
        
        \Large ID 05229 & \Large ID 06176 \\
        \raisebox{-.5\height}{\includegraphics[width=0.40 \textwidth, trim = 25 0 550 25, clip]{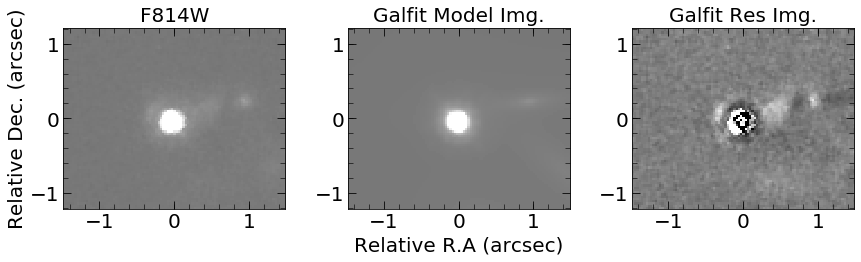}} & \raisebox{-.5\height}{\includegraphics[width=0.40 \textwidth, trim = 25 0 550 25, clip]{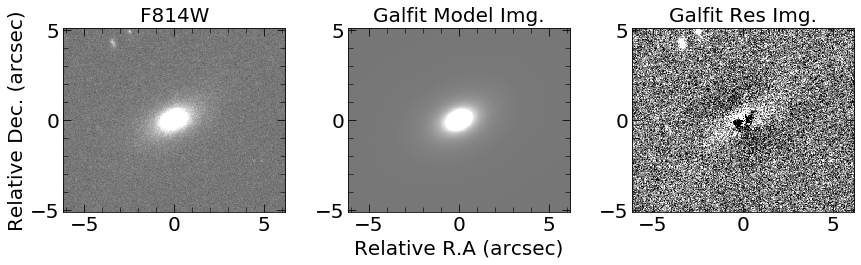}} \\  
    \end{tabular}
    \end{center}
     \caption{Hubble Space Telescope F814W images for galaxies with discrepant classification. The OTELO ID 00026 galaxy image is not available from the AEGIS HST-ACS image archive \citep{davis07}. The left axis corresponds to relative declination in arcsec. The bottom axis indicates the relative right ascension in arcsec.}
    \label{fig:images}
\end{figure*}
\end{center}


\begin{center}
\begin{figure*}[!h]
    \centering
    \begin{center}
    \begin{tabular}{ c c }
        \includegraphics[width=0.4 \textwidth]{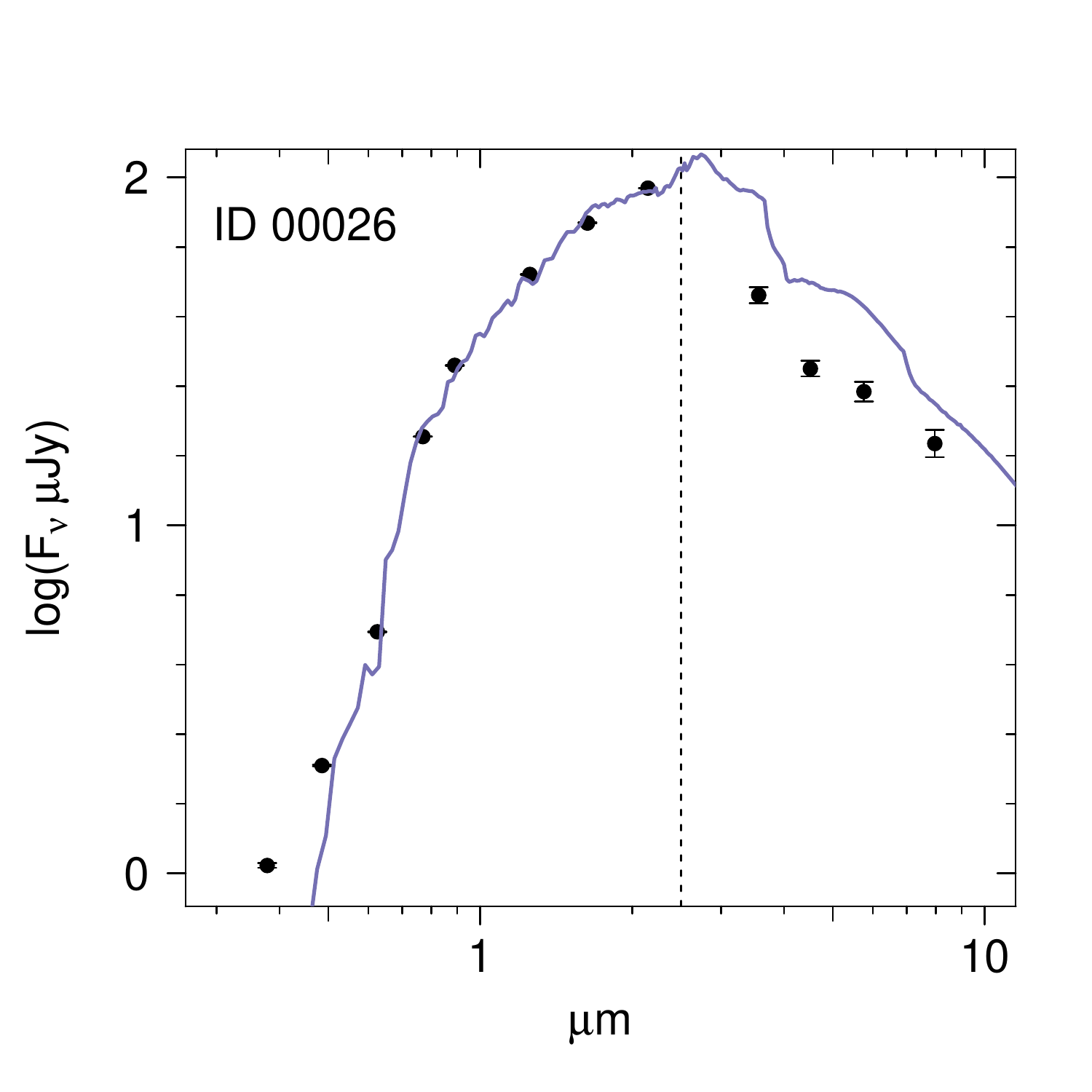} & \includegraphics[width=0.4 \textwidth]{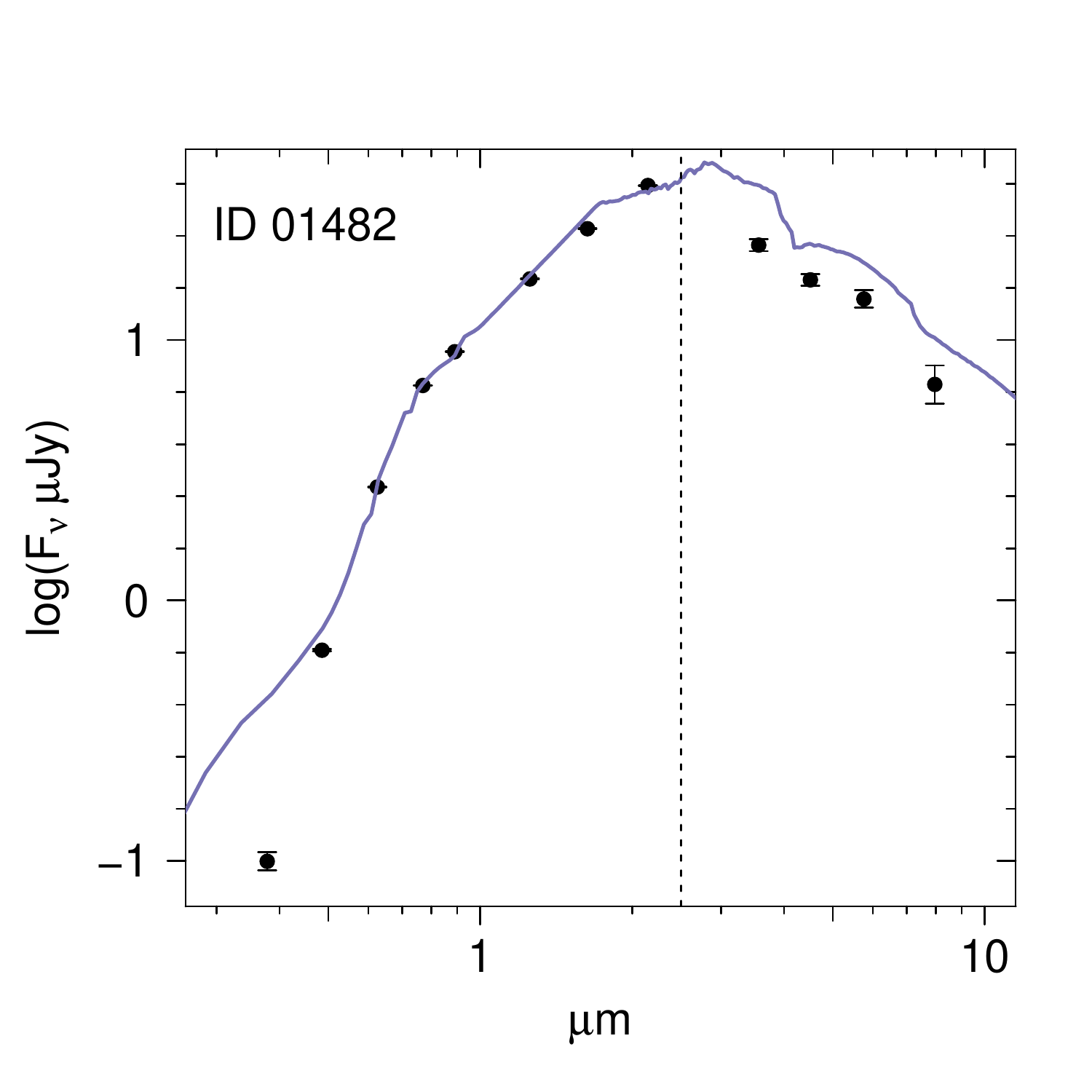} \\
        \includegraphics[width=0.4 \textwidth]{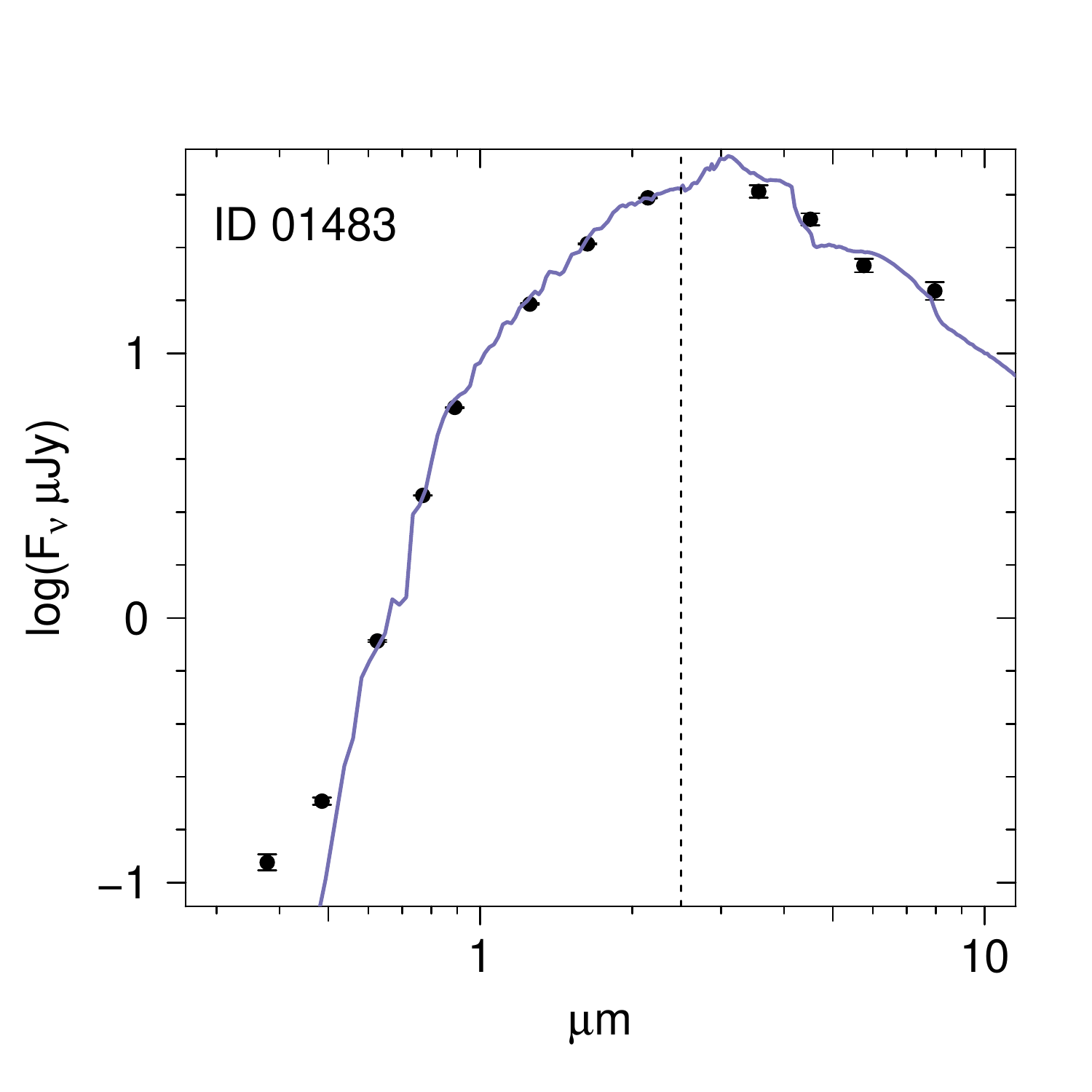} & \includegraphics[width=0.4 \textwidth]{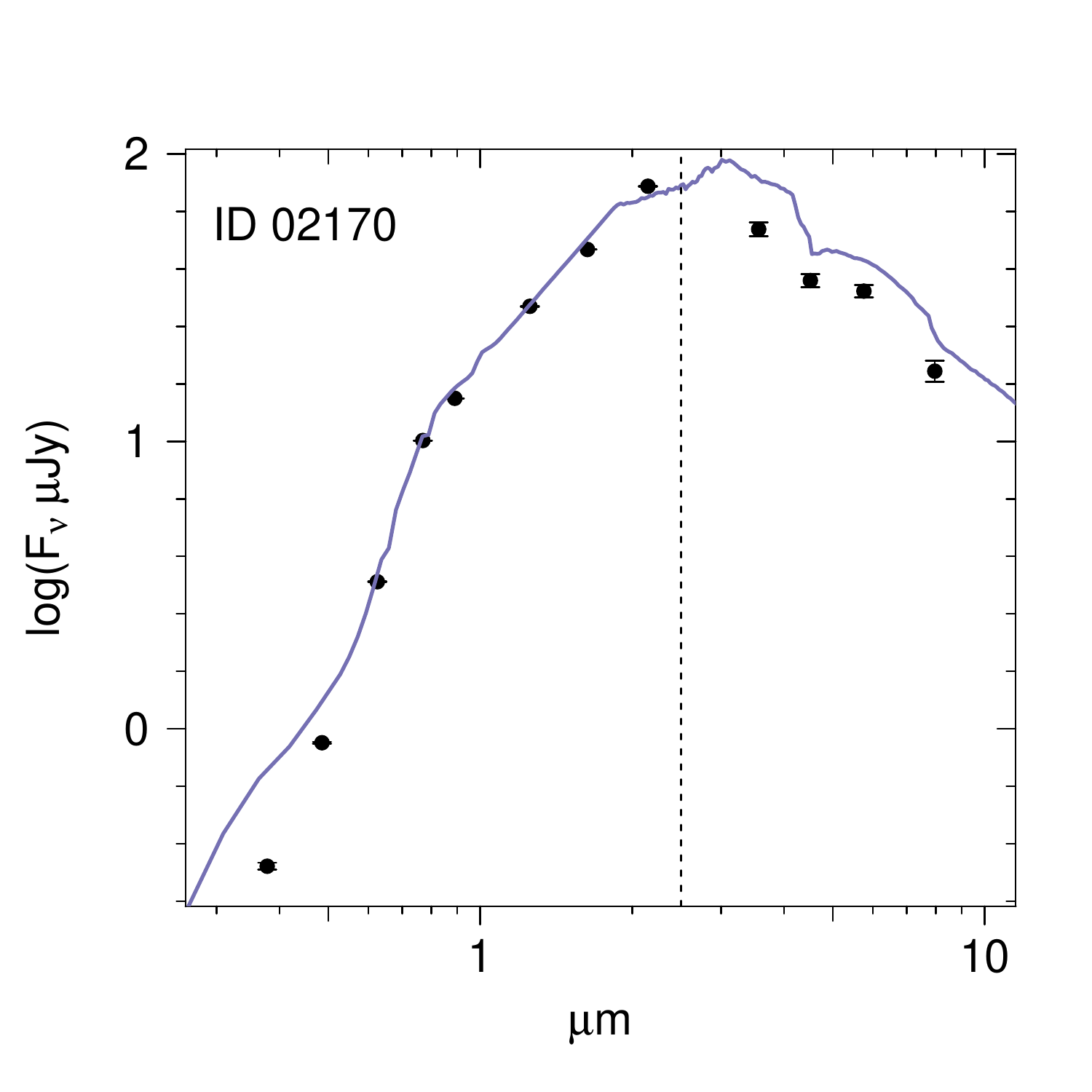} \\ 
        \includegraphics[width=0.4 \textwidth]{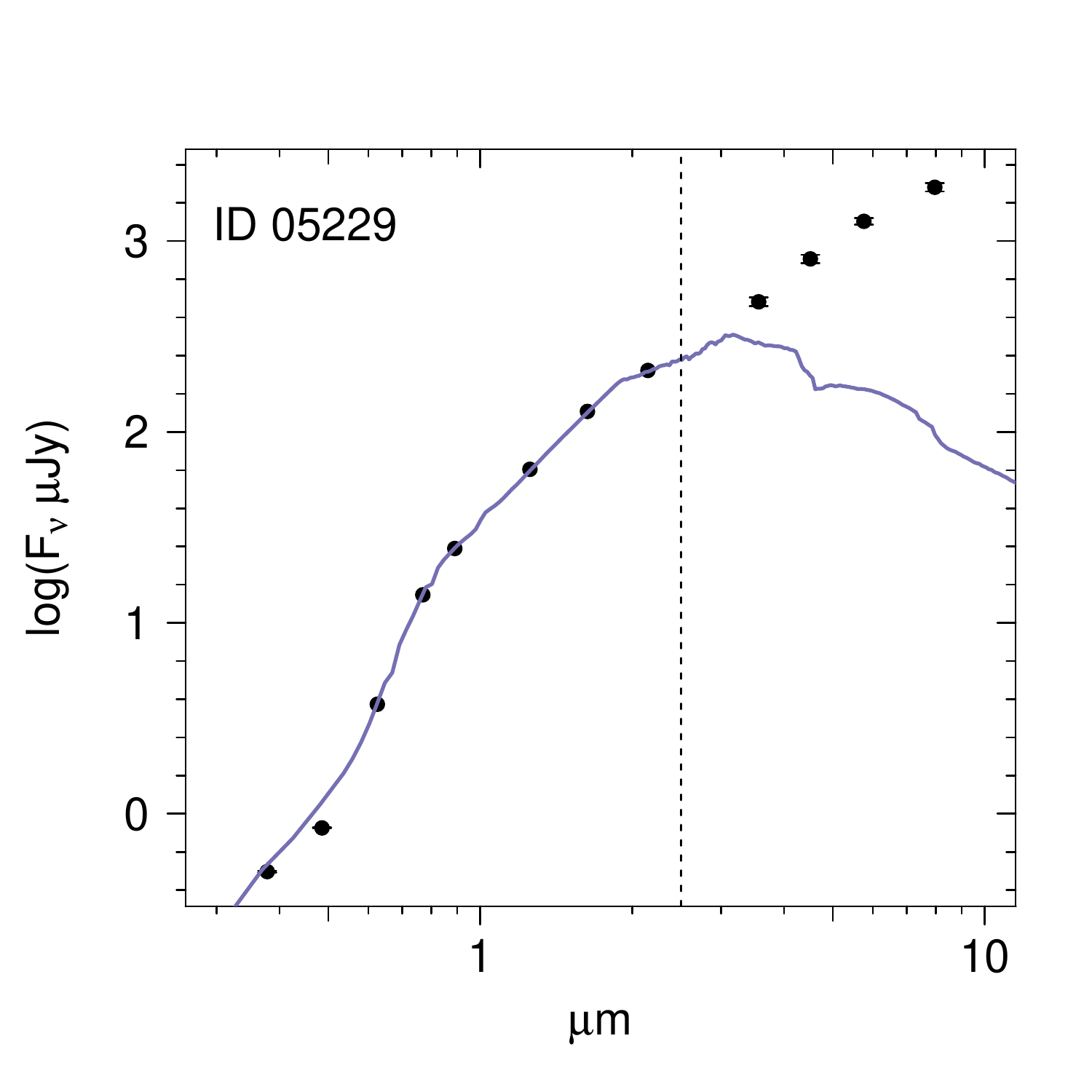} & \includegraphics[width=0.4 \textwidth]{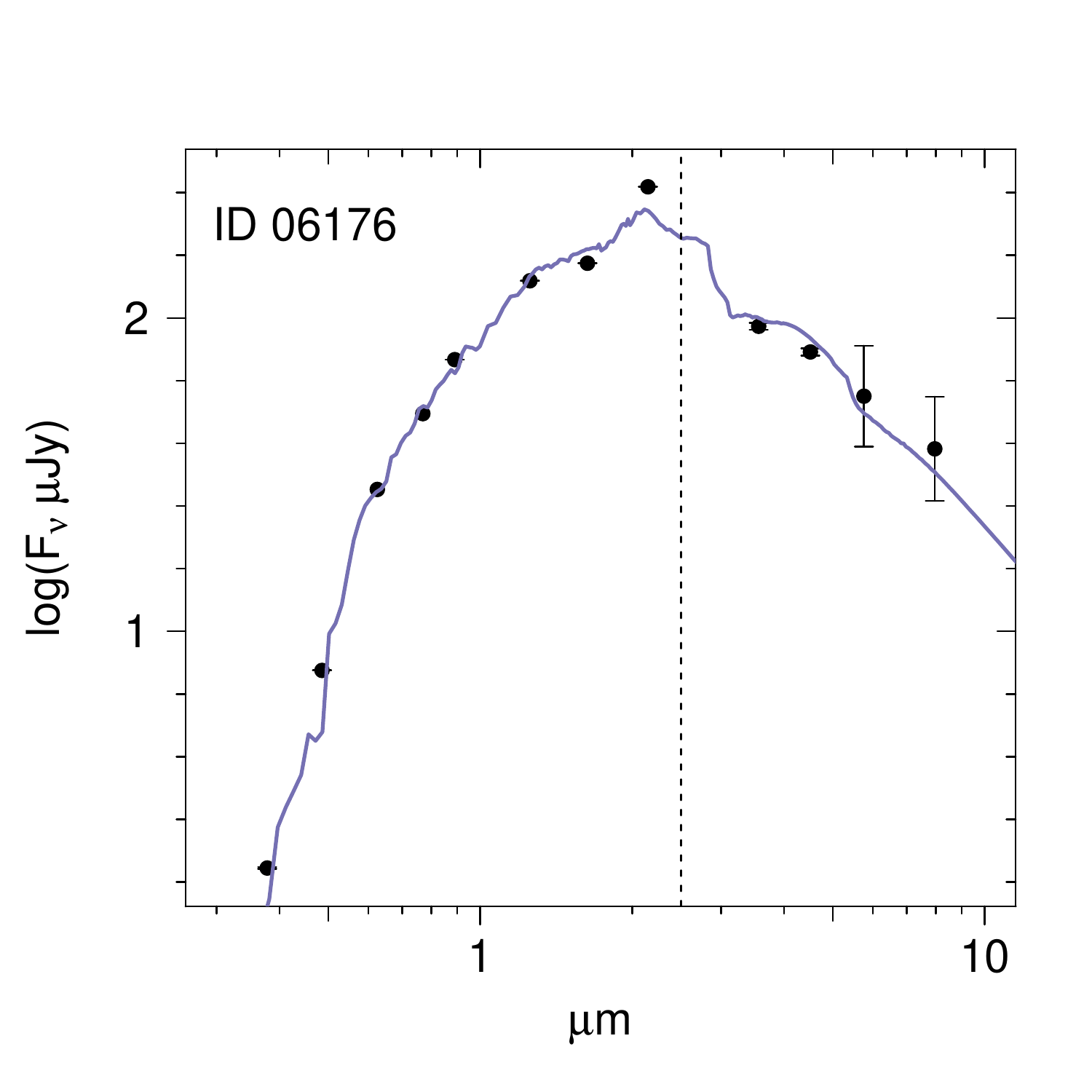}\\  
    \end{tabular}
    \end{center}
    \caption{Spectral energy distributions for galaxies with discrepant classification.  The solid blue line indicates the best-fitting OTELO template.}
    \label{fig:seds}
\end{figure*}
\end{center}


Table~\ref{tab:misclassified} presents the results for the objects with discrepant classifications.   Column 1 in this table shows the OTELO identifier. Columns 2, 3, and 4 indicate the visual classification performed by the authors, OTELO, and the neural network classifications.  Finally, columns 5, 6, and 7 show the spectroscopic redshift, the OTELO, and the neural network estimates.  OTELO employs 4 \citet{coleman80} Hubble-type and 6 \citet{kinney96} starburst templates for the LePhare tool SED fitting.  The Hubble-type templates used are, from earlier to later types, ellipticals (E), spirals (Sbc and Scd), and irregular (Im) galaxies.  We note that all the discrepant classifications were a better fit with the Hubble templates E and Sbc.

Figure~\ref{fig:images} shows the HST Advanced Camera Survey (ACS) F814W images of these misclassified galaxies, but the image for ID~00026 is not available.  The galaxy SEDs are presented in Fig.~\ref{fig:seds}.  Galaxy ID~05229 is actually a mixed group of unresolved galaxies in the GTC-OTELO low-resolution image.  This object is a bright infrared and X-ray source \citep[source irx-8]{georgakakis10}, and OTELO galaxy templates poorly fit the observed SED, probably because the dominant object in the image may be an obscured AGN \citep[source CFRS-14.1157]{zheng04}.  The spectral redshift reported by \citet{matthews13} is 1.14825, but \citet{zheng04} find a redshift of 1.0106 that is in better agreement with both the SED  fitting (0.85) and the neural network prediction (0.98).  The other four galaxies are all visually classified as LT.  Briefly, the visual and OTELO's classifications agree for galaxies ID~001482 and ID~02170, but not for galaxies ID~01483 and ID~06176, for which the visual matches the DNN classification.

\subsubsection{Receiver Operating Characteristic Curve}

\begin{figure}[t]
  \centering
  \includegraphics[width=\columnwidth, trim=0 0 0 25, clip]{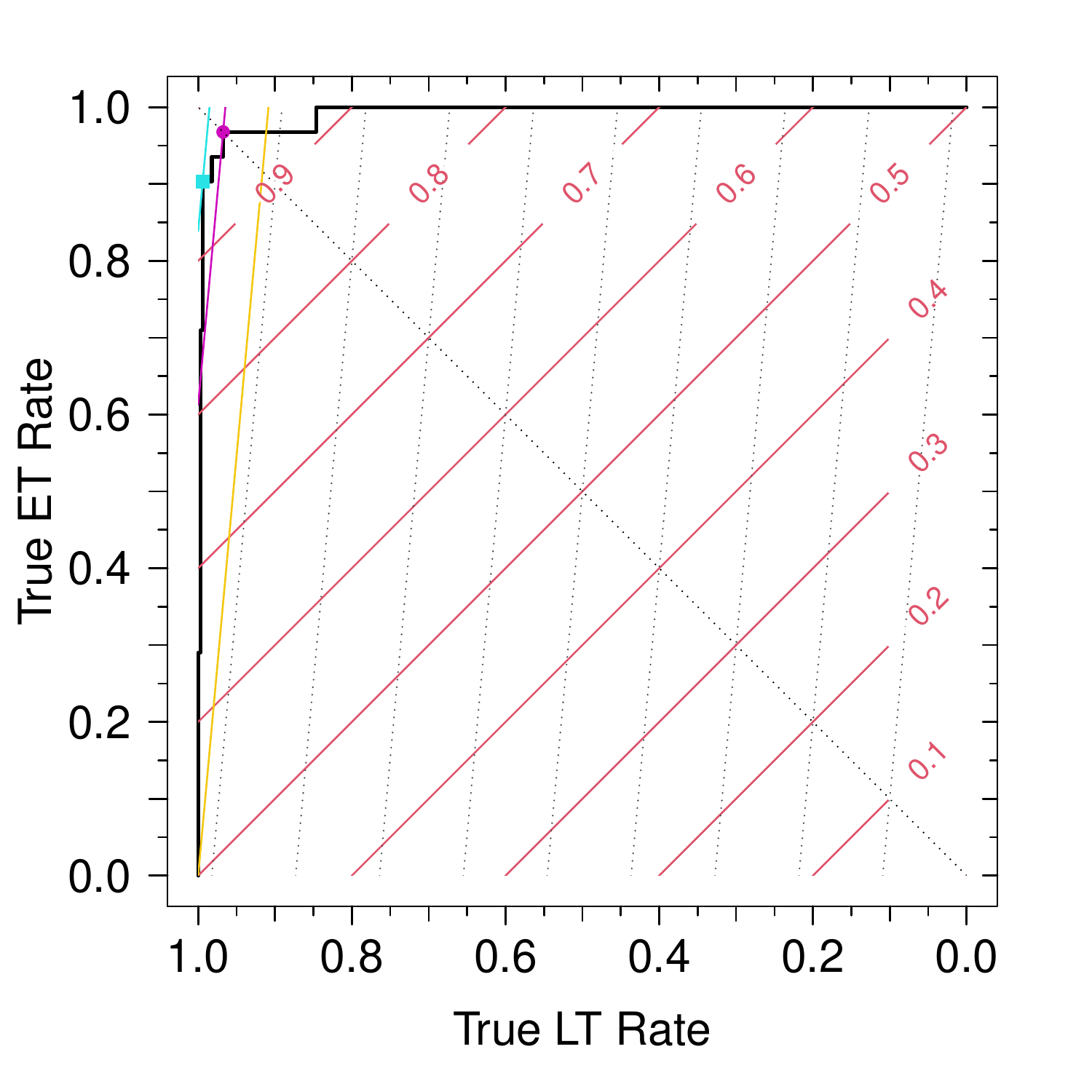}
      \caption{Receiver Operating Characteristic (ROC) curve for galaxy classification.  For clarity, the \emph{specificity} and \emph{sensitivity} axes, common in statistical studies, have been renamed as \emph{True LT Rate} and \emph{True ET Rate}, respectively.  The stepped black line is the ROC curve obtained with the neural network.  The ascending diagonal red lines show the iso-accuracy edges for a balanced sample of ET and LT galaxies.  Similarly, the gray dotted lines represent the iso-accuracy fringes for our imbalanced sample.  Iso-accuracy lines of the same value intersect at the descending diagonal black dotted line.  The orange iso-accuracy line passing through coordinates (1,0) corresponds to the baseline.  The cyan square mark indicates the True LT and ET Rates obtained through the maximum accuracy criterium (\tltr\ and \tetr, respectively).  The magenta circle indicates the rates obtained from Youden's $J$ statistic (0.97 for both the True LT and ET Rates).    
      }\label{fig:roc}
  \end{figure}

Figure~\ref{fig:roc} shows the \emph{Receiver Operating Characteristic} (ROC) curve for the neural network classification.  There are two enhanced points in this figure.  The cyan square mark corresponds to the probability classification threshold $p_{th} = 0.5$, and the magenta circle to Youden's $J$ statistic \citep{youden50}:
$$J = \max(\text{True ET Rate} + \text{True LT Rate} - 1).$$
By combining True ET and LT ratios, Youden's $J$ statistic is the likelihood of correctly ascribing an ET galaxy to the ET class (True ET Rate) versus wrongly ascribing to the ET class galaxies that are LT (1 - True LT Rate).  Thus, the $J$ statistic may have values between 0 and 1, which stand for completely incorrect and perfect classifications, respectively.  In our case, the coordinates of $J$ on the ROC curve are (0.97, 0.97), and the corresponding discriminant threshold probability is $p_J = 0.897$.  Table~\ref{tab:youden} presents the contingencies for the $J$ statistic, and Table~\ref{tab:classification} shows the accuracy, True ET, and True LT rates proportions for the classification baseline, the default threshold ($p_{def} = 0.5$), and Youden $J$ statistic.


We note that classification algorithms do not need to yield precise, population-based probabilities but sample-based scores to discriminate between different categories accurately. Therefore, thresholds usually are not calibrated, and thus they are not proper probabilities \citep{fawcett2004}.  There are several techniques to calibrate machine learning classification scores, being the Platt Scaling \citep{platt1999} and the Isotonic Regression \citep{zadrozny2002}  the most common.  The interested reader may consult \citet{guo2017} for an extensive review of calibration methods for neural networks.

%

Because the maximum accuracy corresponds to the fixed threshold score value of 0.5, accuracy may be a misleading classification metric for imbalanced datasets.  The machine learning algorithms tend to assign elements into the majority class, while the high accuracy produces the false sense that the model is better than it is. There is no catch-all receipt for all the classification problems. The method used depends on the researcher's interest, and it constitutes a subject of recent investigation \citep[\eg][]{flach2003, furnkranz2005, armano2015, zou16, boughorbel2017, haixiang2017, raghuwanshi2020}.

ROC curves' fundamental property is that they are invariant to the prevalence of the dataset classes \citep{metz78}.  Thus, in our case, this property is a consequence of the fact that the True ET Ratio depends only on the proportion of ET galaxies recovered in the sample, and the same is applicable for the True LT Ratio and the proportion of LT galaxies.  If our dataset had shown a perfect balance between ET and LT galaxies, the ROC curve would remain essentially the same but for sampling variance.  On the contrary, classification accuracy strongly depends on the prevalence of the different classes:
$$A = \frac{R_{TET} N_{ET} + R_{TLT} N_{LT}}{N_{ET} + N_{LT}} = \frac{R_{TET} + p_c R_{TLT}}{p_c + 1},$$
where $A$ is the accuracy; $R_{TET}$ and $R_{TLT}$ the True ET and LT ratios, respectively; $N_{ET}$ and $N_{LT}$ the respective total number of ET and LT galaxies; and $p_c = N_{LT} / N_{ET}$ the prevalence class ratio.

Figure~\ref{fig:roc} shows the iso-accuracy lines for both a theoretically balanced dataset (red lines) and our imbalanced sample (in gray).  The iso-accuracy lines for the balanced data intersect the ROC curve at two points with inverted but approximately equal True LT Rate and True ET Rate values.  However, the iso-accuracy lines for the imbalanced data intersect the ROC curve at significantly different values.  For example, the baseline iso-accuracy edge ---orange line--- intersects the curve at (1,0) and approximately (0.91, 0.96).  The high dependence of the True ET Rate (but not the True LT Rate) on the accuracy is a non-desired consequence of our highly imbalanced data.

The difference between the True LT and ET rates associated with our network's maximum accuracy at the probability threshold of 0.5 reveals that we are misclassifying a larger proportion of ET than LT galaxies (see Fig.~\ref{fig:roc}).  For maximum accuracy, the number of misclassified objects is similar for both galaxy types (in this case, it is the same by chance), as shown in Table~\ref{tab:contingency}.  However, if we apply our trained network and the 0.5 thresholds to another sample consisting of a different proportion of ET and LT galaxies, we will obtain similar proportions of misclassified objects, but we will not attain the same accuracy the model achieves in the current sample.  It is well known that ET and LT galaxies do not distribute the same in different environments \citep[\eg][]{calvi12} and present opposite distribution gradients in clusters \citep[\eg][]{doi95}.

If the accuracy varies depending on the sample characteristics, it is not the best choice for deciding the probability threshold for classification.  Youden's $J$ statistic provides well-balanced True LT and ET rates that will deliver comparable results no matter the class prevalences in the sample.


\subsection{Photometric redshifts}

\citet{bongiovanni19} fitted LePhare templates to multifrequency broadband data to estimate photometric redshifts in the OTELO catalog, achieving a precision of $\Delta z \sim 0.2$, which is typical for this kind of estimates.  Previous studies have shown that neural networks often provide better assessments than SED fitting \citep{firth03, tagliaferri03, singal11, brescia14, eriksen20}.  For galaxies at $z \lesssim 0.4$ errors are $\Delta z_{phot} \simeq 0.02$ \citep{firth03, tagliaferri03, brescia14} even using a small database of 155 galaxies for training and 44 for testing \citep{vanzella04}.  Errors increase as the samples extend to redshifts $z \gtrsim 1$ \citep{firth03, tagliaferri03} reaching values of $\Delta z \simeq 0.1$ \citep{vanzella04}, and decrease as the number of available photometric bands gets larger, as for the 40 narrow bands sample of \citet{eriksen20}.  Besides the photometric data, some authors have provided morphological information to improve the neural network's redshift estimates, with inconclusive results.  Thus, \citet{singal11} and \citet{wilson20} report that including, in addition to the photometry, several shape parameters such as the concentration, Sérsic index, the effective radius, or the ellipticity, do not produce any improvement in the redshift estimate.  In contrast, \citet{tagliaferri03} include photometry, surface brightnesses, and Petrosian radii and fluxes, reducing the errors by a factor of about 30\%.  More recently, \citet{menou19} presents a non-sequential network with two inputs and one output, a convolutional branch to analyze images, and a branch composed of a dense set of layers to deal with the photometric data, reporting an improvement in the redshift estimates.
For our redshift analysis, we take into account the probability yielded by the Classification module.

\begin{table}[t!]
    \centering
    \caption{Contingency table using Youden's J statistic threshold.}\label{tab:youden} 
    \begin{tabular}{rrrr}
        \hline\hline
                 & ET\so & LT\so & Sum \\ 
        \hline
        ET\sdnn  & 30    &  11   &  41 \\ 
        LT\sdnn  &  1    & 328   & 329 \\ 
        Sum      & 31    & 339   & 370 \\ 
        \hline
    \end{tabular}
\end{table}

\begin{table}
    \centering
    \caption{Classification statistics}\label{tab:classification} 
\begingroup
    \setlength{\tabcolsep}{3pt} 
    \begin{tabular}{lr@{$\,\pm\,$}lr@{$\,\pm\,$}lr@{$\,\pm\,$}l}
        \hline\hline
        \multicolumn{1}{c}{Classifier}      & \multicolumn{2}{c}{Accuracy} & \multicolumn{2}{c}{True ET Rate} & \multicolumn{2}{c}{True LT Rate} \\ 
        \hline
        Baseline        & 0.92 & 0.02   & \multicolumn{2}{c}{0\phantom{000}} & \multicolumn{2}{c}{1\phantom{0}}  \\ 
        Default Th.     & 0.984 & 0.007    & 0.90 & 0.06   & 0.991 & 0.006 \\ 
        Youden's J      & 0.97 & 0.01    & 0.97 & 0.04   & 0.97 & 0.01 \\        
        \hline
    \end{tabular}
\endgroup
\end{table}

\begin{figure}[t]
  \centering
  \includegraphics[width=\columnwidth, trim=10 15 25 25, clip]{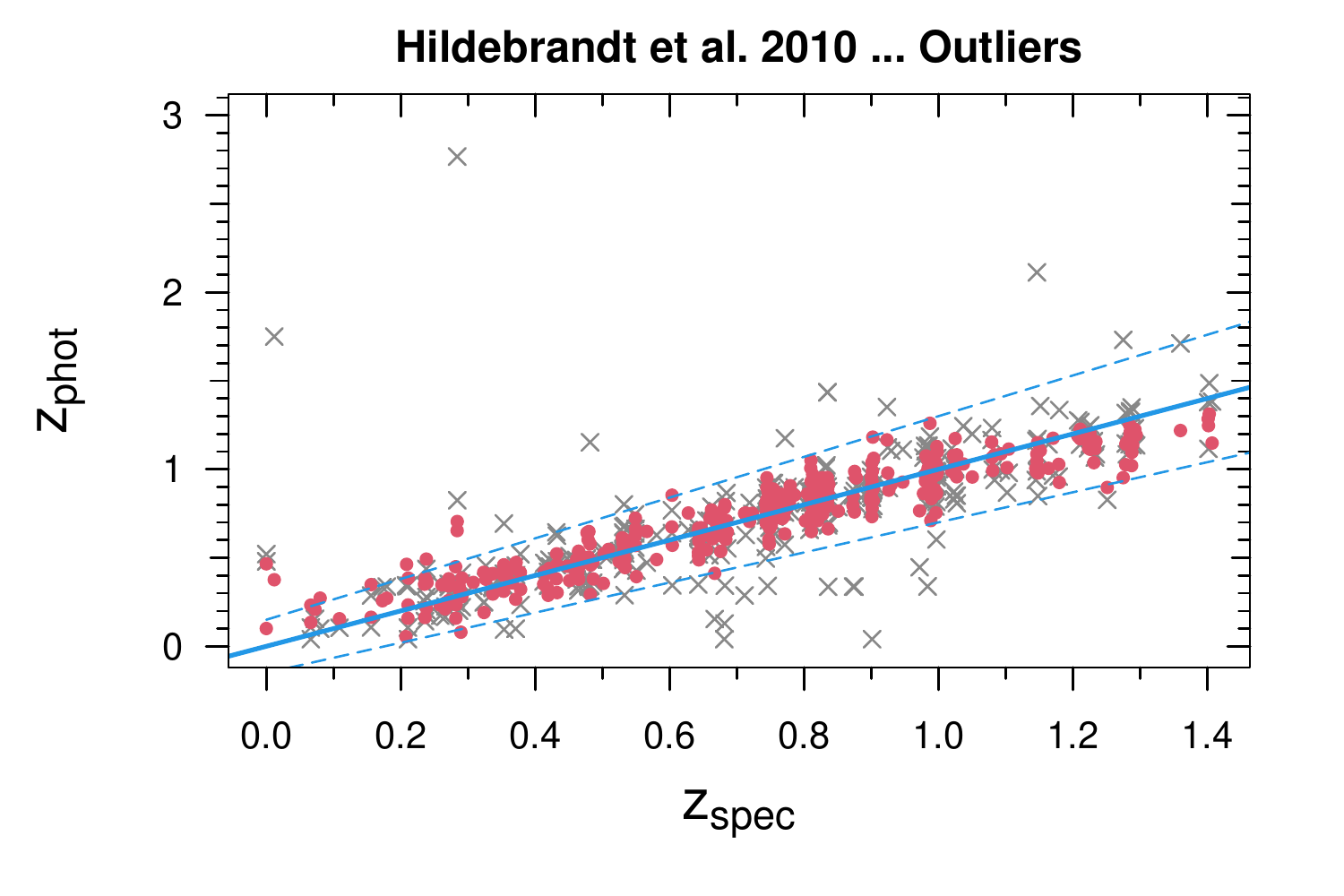}
      \caption{Photometric vs. spectroscopic redshift comparison. Gray crosses indicate the OTELO template fitting photometric redshifts and filled red circles the neural network estimates.  The regression lines for the different fits —not plotted— are barely distinguishable from the perfect correlation between photometric and spectroscopic redshifts (blue-solid thick line).  The blue-dashed thin lines show the \citet{hildebrandt10} outlier limits.} \label{fig:z}
  \end{figure}

Figure~\ref{fig:z} shows the photometric vs.\ spectral redshifts for the galaxies in our sample, the linear regression fit lines, and the outlier limits  $\delta_z > 0.15$, with $\delta_z$ the precision as provided by \citet{hildebrandt10}:

$$\delta_z = \frac{\mid z_{ph} - z_{sp} \mid}{1 + z_{sp}},$$  

\noindent
where $z_{ph}$ and $z_{sp}$ are the photometric (either OTELO or DNN), and the spectral redshifts, respectively.  The total number of outliers is 34 for the OTELO photometric redshifts and 13 for the neural network,  corresponding to $(9 \pm 2)\%$ and $(4 \pm 1)\%$ of the \ngal galaxies in our sample, respectively. 
A simple inspection of this figure reveals that the neural network estimates strongly mitigates the amount and magnitude of the catastrophic redshift outliers produced by the template fitting.  Table~\ref{tab:z} shows that the residual means are indistinguishable from zero for both estimates, but the residual standard deviation attained with the neural network (0.11) reduces in a factor of two the dispersion achieved through the template-based fitting (0.23).  

\begin{figure}
    \centering
    \includegraphics[width=\columnwidth, trim=0 10 0 0, clip]{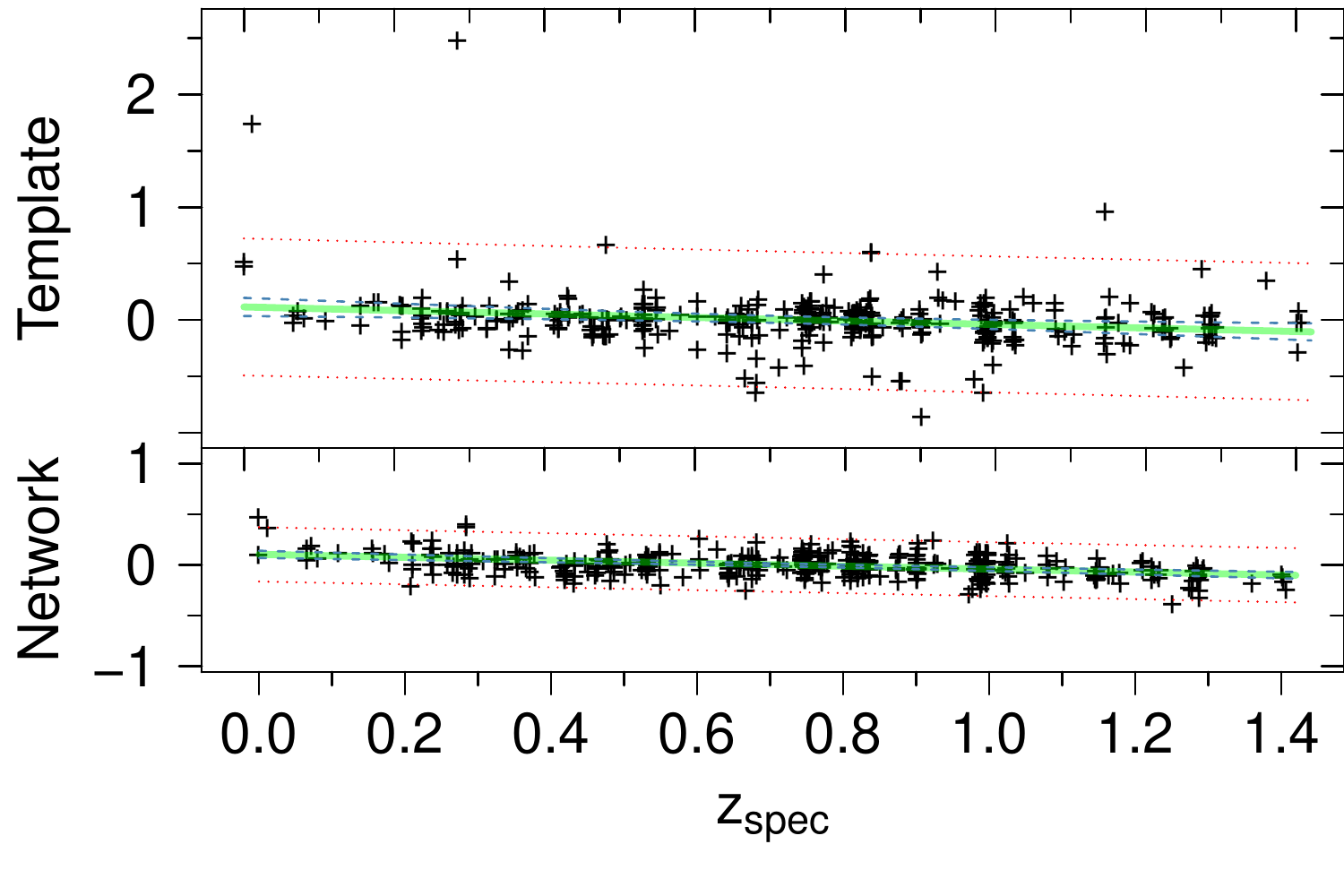}
    \caption{Photometric redshift residuals. Obtained from LePhare template fittings (upper panel) and neural network predictions (lower panel). The residuals are plotted as plus signs, the regression fit as a thick green solid line. Dashed blue lines indicate 99\% confidence bands and pointed red lines the 99\% prediction bands.}
    \label{fig:residuals}
\end{figure}

Figure~\ref{fig:residuals} shows the residuals for the template and neural network estimates as a function of the redshift.  The residuals show small negative slopes, indicating a minor lack of adjustment for both models, overestimating redshifts $z \lesssim 0.2$ and underestimating redshifts $z \gtrsim 1.3$.  High-redshift galaxies are prone to be dimmer, and thus the photometric errors are likely to increase with redshift; this may explain, at least partially, the small misadjustment for galaxies at $z \gtrsim 1.3$.  However, there are two additional causes in the neural network that may contribute to this lack of fitting.  The first one is that neural networks are interpolating approximation functions; at low-redshifts, the predicted values must be larger than zero, and at high-redshifts less than 1.4 because these are the constraints in the training sample range of redshifts. The second cause is associated with the small number of galaxies at low and high-redshifts shown in Fig.~\ref{fig:distribution}.  Thus, because of the insufficient number of subjects in the training sample, the network tends to push the redshift estimates towards the most populated redshift areas, increasing low-redshift and decreasing high-redshift estimates.

The standard deviation of our sample residuals, $\sigma_z = 0.11$, improves the redshift accuracy obtain \citet{vanzella04} for their spectroscopic redshift sample ($\sigma_z = 0.14$) using 150 Hubble Deep Field North galaxies for training and 34 Hubble Deep Field South objects at $z < 2$ for validation.  Also the median precision $\tilde{\delta}_z =0.040 \pm 0.002$ improves \citeauthor{vanzella04} $\tilde{\delta}_z = 0.06$ results significantly.  Comparing our results with other studies is less straightforward because of the large number of galaxies involved and the redshift's reduced range.  For example, \citet{brescia14} used a sample of \numprint{497339} Sloan galaxies at $z < 1$, of which 90\% were at $z < 0.345$ and with a distribution peak at $z \approx 0.1$.  In contrast, only 13\% of our sample is located at $z < 0.345$, and the distribution peaks at $z \approx 0.8$ (see Fig.~\ref{fig:distribution}).

\subsection{Effect of the classification on the redshift estimate}

The galaxy colors are related to both the redshift and the morphology. However, the contradictory results obtained by other researchers for and against including morphological data to increase the redshift estimate accuracy \citep{tagliaferri03, singal11, menou19, wilson20} suggest that the morphological properties provide limited information to attain this goal.  Therefore, we used bootstrap resampling to analyze the effect of the galaxy classification probabilities passed on the redshift module in our neural network model.  We considered that dropouts suppress a fraction of the layer outputs but do not change neurons' weights, which are continually updated during training.  Therefore, the concatenate layer in our model has 513 inputs and their corresponding weights; 512 of these inputs correspond to the outputs from the Base module, regardless of this module dropout, and only one linked to the Classification module output. In addition, the concatenate layer has its own weights.

\begin{table}
    \centering
    \caption{Redshift residuals.}\label{tab:z}
    \begin{tabular}{lr@{$\,\pm\,$}lr@{$\,\pm\,$}l}
        \hline\hline
                 & \multicolumn{2}{c}{OTELO} & \multicolumn{2}{c}{DNN} \\ 
        \hline
        Mean      &  0.00 & 0.02   &   0.00 & 0.03 \\ 
        Sd. Dev.  &  0.23 & 0.06   &   0.11 & 0.02 \\ 
        Intercept &  0.11 & 0.04   &   0.10 & 0.02 \\
        Slope     & -0.16 & 0.04.  &  -0.15 & 0.02 \\
   \hline
\end{tabular}
\end{table}

Because of the indeterministic nature of the neural networks, hidden layer neurons estimate different relations in each training run.  Consequently, the weights will differ from one bootstrap run to another, and only a few in each run will show an absolute value large enough to contribute to the result significantly.  Thus, monitoring a single random weight during different runs will reproduce the layer weight distribution.  The only exception may be the weight that is always associated with the output of the classification module if this factor has some effect on the concatenate layer outcome.  In this case,  we expect an absolute larger weight and possibly a different weight distribution shape.

\begin{figure}
    \centering
    \includegraphics[width=\columnwidth, trim=0 10 0 30, clip]{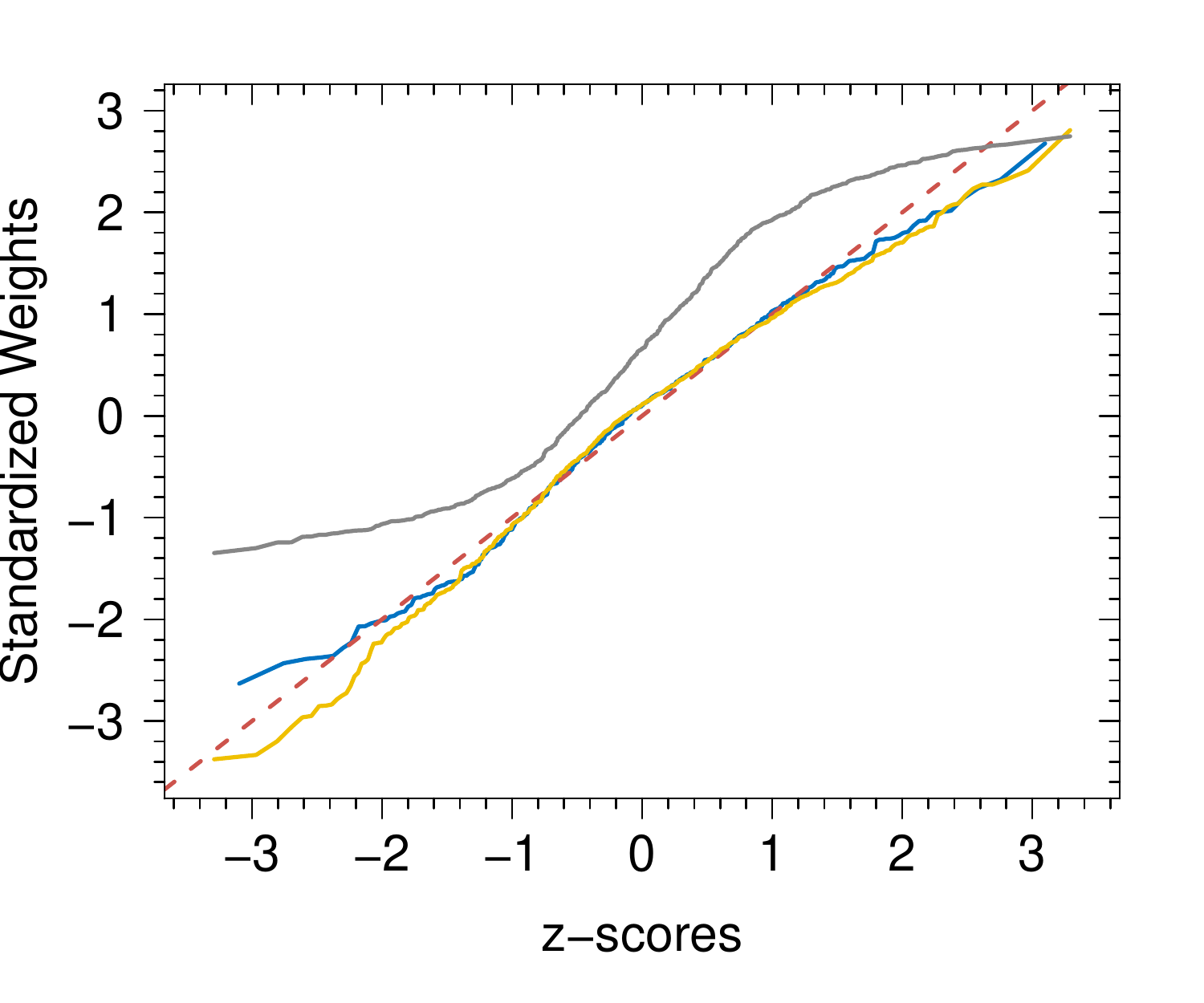}
    \caption{Q-Q plot of the weights of the concatenate layer.  The solid blue line shows the standardized distribution of the 513 concatenate layer weights obtained in a single training.  The solid gray line displays the standardized distribution of the weights corresponding to the neuron linked to the output of the Classification module, obtained from 1000 bootstrap runs.  The blue line shows the standardized distribution of the bootstrap runs for a random neuron weight. The red dashed line indicates the expected relation for a standard normal distribution.}
    \label{fig:concat}
\end{figure}

We have performed 1000 bootstrap training runs with the final architecture to analyze the weight distribution of the concatenate layer and the neuron associated with the classification output in this layer.  We used all the galaxies from our sample as the training set for these runs (neither validation nor test sets were necessary for this task).  Figure~\ref{fig:concat} shows the quantile-quantile (Q-Q) plots for the weights of the whole concatenate layer, a randomly chosen neuron, and the neuron linked to the classification module output.  In the case of the whole concatenate layer, we extracted the weights from a single run; the shape of the distribution does not change from run to run.  For easy comparison with the other distributions presented in the figure, we have standardized them all by subtracting the mean and dividing by the standard deviation of the layer weights; for individual neurons, we standardized each weight using the layer distribution statistics corresponding to each bootstrap run. We note that the distribution of the concatenate layer weights (solid blue line in Fig.~\ref{fig:concat}) presents a normal-like distribution (dashed red line).  As expected, the distribution of the bootstrap runs of a randomly selected neuron in this layer (solid yellow line) closely mirrors the distribution of the whole layer.  However, the distribution of the bootstrap runs of the neuron associated with the output of the classification layer (gray solid line) is quite different.  It contrasts with the normal distribution due to its upwards shift that indicates a central value different from zero (median of $0.66 \pm 0.05$), and its appearance evinced by the S-shaped line that corresponds to a thin tailed or platykurtic distribution with kurtosis of $1.81 \pm 0.04$.  Platykurtic distributions and medians are robust to the presence of outliers, and thus the different behavior of the classification neuron cannot be ascribed to a few extreme neuron weights.  For comparison, the random neuron weights show statistics more similar to the normal distribution, with a median of $0.11 \pm 0.04$ and kurtosis of $3.1 \pm 0.2$.

In total, 41\% and 13\% of the classification neuron weights are greater than 1$\sigma$ and 2$\sigma$, respectively.  These results contrast with those for the random neuron and the concatenate layer weights, both with 15\% and 1\% performance at these $sigma$ values. However, even if the differences are notable, there is a rather large probability that the classification neuron does not significantly affect the redshift estimate for a single training.  This behavior may be a consequence of the limited training due to our sample's small number of objects.  Regardless of a definitive explanation, whether or not the weight is large enough to influence the redshift estimate may explain the disagreement between the results obtained by the different groups that have included morphological parameters in their neural networks for photometric estimates of the redshift.


Our analysis using a non-sequential neural network model has shown that it is feasible to incorporate morphological classification to estimate galaxies' redshifts.  It is worth noting that we have used galaxy photometry \emph{as it is}, without extinctions and k-corrections that would incorporate theoretical models into a pure empirical methodology.
The number of OTELO galaxies with spectroscopically measured redshifts (\ngal) is rather small for neural network training.  However, we have obtained classification and photometric redshift accuracies that compete with and improve those obtained through SED template fitting.
The redshift range of the OTELO spectroscopic sample ends at $z \simeq 1.4$, just at the beginning of the redshift desert ($1.4 < z < 3$).  Future work should cover this and higher redshift regions where obtaining bulk spectroscopic measurements has been challenging until the recent development of near-infrared multi-object spectrographs \citep[\eg MOSFIRE][]{mclean10, mclean12}.  

These results show that non-sequential neural networks can produce several outcomes simultaneously.  By sharing repeated calculations between the fitting procedures, the computing and training total times reduce by a factor that depends on the complexity of the procedure and shared intermediate byproducts.  Moreover, transferring results between procedures ensures the internal consistency of the predictions.

\section{Conclusions}

The first methods for automatically analyzing galaxy surveys relied on the SED fitting of observed or simulated spectra.  These methods allow simultaneous morphological classification and photometric redshift estimates of the sources at the expense of many catastrophic mismatches.  Machine-learning-based models ---support vector machines, random forest, and neural networks--- are becoming more and more popular for analyzing survey data, and they have shown to be more robust than SED fitting for redshift estimates. The interest in neural networks has increased spectacularly during the last five years, fuelled by advances in hardware and tensor-oriented libraries.  Some advantages over other machine learning techniques are that neural networks efficiently deal with missing data and feed galaxy images directly into the model.
However, machine learning morphological classification and photometric redshift estimates have been computed independently with different models.  This split in the machine learning methods may produce inconsistencies due to the tendency to level low-redshift ET and high-redshift LT galaxy colors.

This study tested a non-sequential neural network architecture for simultaneous galaxy morphological SED-based classification and photometric redshift estimates, using optical and near-infrared photometry of the 370 sources with spectroscopic redshift listed in the OTELO survey.  These photometric data were not subjected to extinction or k-corrections.
We found that our non-sequential neural network provides accurate classification and precise photometric redshift estimates. Besides, sharing the classification results tends to improve the redshift estimates and the congruence of the predictions. 
Our results demonstrate the non-sequential neural networks' capabilities to deal with several problems simultaneously and share intermediate results between the network modules.  Furthermore, the analysis of the neuron weights shows that including the morphological classification prediction often improves the photometric redshift estimate.

We have used a ROC curve and the Youden J statistic to obtain a classification threshold that provides a better equilibrium between misclassified galaxy types and reduces the problem of imbalanced samples.
Besides, our neural network reduces the mean error of the photometric redshift estimates by a factor of two relative to the SED template fitting.

This study strongly suggests that the benefits gained by simultaneously treating morphological classification and photometric redshift estimates may also address other prediction problems across a wide range of astronomical problems, particularly when consistency between the results is required.
Most notably, this is the first study to our knowledge that provides simultaneous and consistent predictions using a single deep learning methodology, a non-sequential neural network model.


The OTELO sample has some limitations that are worth noting.  The first one is the low number of sources with measured spectral redshifts; machine learning methods perform better with many available training data.  Besides, the scarcity of spectral redshift measurements for low ($z < 0.2$) and high-redshift ($z > 1.3$) sources limits the precision of the estimates at these redshift ranges.  The last limitation is the redshift cut-off at $z \approx 1.4$ that restricts the range of our study.

Currently, the OTELO project is conducting spectroscopical monitoring for about two hundred new objects using the OSIRIS MOS instrument attached at the GTC.  These observations will add certainty about the nature and the redshifts of H$_ {\beta}$, H$_{\alpha}$, [OIII], and HeII emitter candidates, as well as compact early elliptical galaxies. In addition, the results will be compared with estimates obtained from our network.  

Future work should include a large and redshift balanced sample of galaxies and extend the redshift range to cover all observable sources with modern instruments.  For resolved galaxies, models allowing both images and tabulated data —multi-input non-sequential neural networks— can benefit from multi-band photometry and morphological profile aspects for better classification.  In particular, the photometry of spiral galaxies depends on their orientation, changing the relative contribution of disk and bulge components. Thus,  galaxy classification will benefit from the support of images that may disentangle orientation effects.  

\begin{acknowledgements}
    The authors gratefully thank the anonymous referee for the constructive comments and recommendations, which helped improve the paper's readability and quality.\\
    This work was supported by the project Evolution of Galaxies, of reference AYA2014-58861-C3-1-P and AYA2017-88007-C3-1-P, within the "Programa estatal de fomento de la investigación científica y técnica de excelencia del Plan Estatal de Investigación Científica y Técnica y de Innovación (2013-2016)" of the "Agencia Estatal de Investigación del Ministerio de Ciencia, Innovación y Universidades", and co-financed by the FEDER "Fondo Europeo de Desarrollo Regional".\\
    JAD is grateful for the support from the UNAM-DGAPA-PASPA 2019 program, the UNAM-CIC, the Canary Islands CIE: Tricontinental Atlantic Campus 2017, and the kind hospitality of the IAC. \\
    MP acknowledges financial supports from the Ethiopian Space Science and Technology Institute (ESSTI) under the Ethiopian Ministry of Innovation and Technology (MoIT), and from the Spanish Ministry of Economy and Competitiveness (MINECO) through projects AYA2013-42227-P and AYA2016-76682C3-1-P, and from the Spanish Ministerio de Ciencia e Innovaci\'on - Agencia Estatal de Investigaci\'on through projects PID2019-106027GB-C41 and AYA2016-76682C3-1-P.\\
    APG, MSP and RPM were supported by the PNAYA project: AYA2017--88007--C3--2--P.\\
    JG was supported by the PNAYA project AYA2018--RTI-096188-B-i00.\\
    MC \& APG are also funded by Spanish State Research Agency grant MDM-2017-0737 (Unidad de Excelencia María de Maeztu CAB).\\
    JIGS receives support through the Proyecto Puente 52.JU25.64661 (2018) funded by Sodercan S.A. and the Universidad de Cantabria, and PGC2018--099705--B--100 funded by the Ministerio de Ciencia, Innovación y Universidades.\\
    EJA acknowledges funding from the State Agency for Research of the Spanish MCIU through the “Center of Excellence Severo Ochoa" award to the Instituto de Astrofísica de Andalucía (SEV-2017-0709) and from grant PGC2018-095049-B-C21.\\
    Based on observations made with the Gran Telescopio Canarias (GTC), installed in the Spanish Observatorio del Roque de los Muchachos of the Instituto de Astrofísica de Canarias, in the island of La Palma. This work is (partly) based on data obtained with the instrument OSIRIS, built by a Consortium led by the Instituto de Astrofísica de Canarias in collaboration with the Instituto de Astronomía of the Universidad Autónoma de México. OSIRIS was funded by GRANTECAN and the National Plan of Astronomy and Astrophysics of the Spanish Government.
\end{acknowledgements}

%
%


\bibliographystyle{aa} 
\bibliography{multioutput} 

\end{document}